\documentclass{WileyMSP-template}
\usepackage[utf8]{inputenc} 
\usepackage{graphicx}
\usepackage{dcolumn}
\usepackage{bm}
\usepackage{amsmath}
\usepackage{cite}
\usepackage{amssymb}
\usepackage{enumitem}
\usepackage{hyperref} 
\usepackage{matlab-prettifier}
\usepackage[user]{zref}

\hypersetup{
    colorlinks=true,
    linkcolor=blue,
    citecolor=blue,
    filecolor=blue,
    urlcolor=blue,
}

\def\e{\begin{equation}}
\def\f{\end{equation}}

\def\.{\cdot}

\def\@#1{_{\rm #1}}

\def\vect#1{\bm{\mathrm{#1}}}
\def\uvect#1{\hat{\bm{\mathrm{#1}}}}
\def\dyad#1{\overline{\overline{#1}}}
\def\tg{{\vert\vert}}

\begin{document}

\pagestyle{fancy}
\rhead{\vspace{0.5cm}}

\title{A General Expression for Homogeneous Metasurface Scattering}

\maketitle


\author{F. S. Cuesta}
\author{K. Achouri*}



\begin{affiliations}
Dr. F. S. Cuesta, Assist. Prof. K. Achouri\\
\'{E}cole Polytechnique F\'{e}d\'{e}rale de Lausanne, Route Cantonale, Lausanne, 1015, Switzerland\\
Email Address: francisco.cuestasoto@epfl.ch, karim.achouri@epfl.ch
\end{affiliations}



\begin{abstract}
The general approach in metasurface design is to find the unit-cell properties required to achieve a given functionality. This is usually done by modeling the metasurface as a combination of surface electric and magnetic polarization densities, whose parameters are determined by solving the generalized sheet transition conditions. This is a time consuming task, as the so-called boundary conditions needs to be solved per-case basis, depending on the source polarization, angle of incidence, and intended functionality. Evermore, the task complexity increases as factors such as different media around the metasurface and more than one illumination scenario are taken into account. 
In this work, we provide a general solution for a uniform metasurface homogenized in terms of susceptibilities. With this model, it is possible to obtain analytical expressions for the specular scattering produced by a metasurface illuminated with arbitrary illumination and angle of incidence. It is expected that the proposed model can ease the analysis and design of metasurfaces, by providing straight-forward expressions which can be simplified by exploiting the unit-cell symmetries.

\end{abstract}


\section{Introduction}\label{sec:intro}

\begin{figure}[h]
\centering
\includegraphics[width=0.45\linewidth]{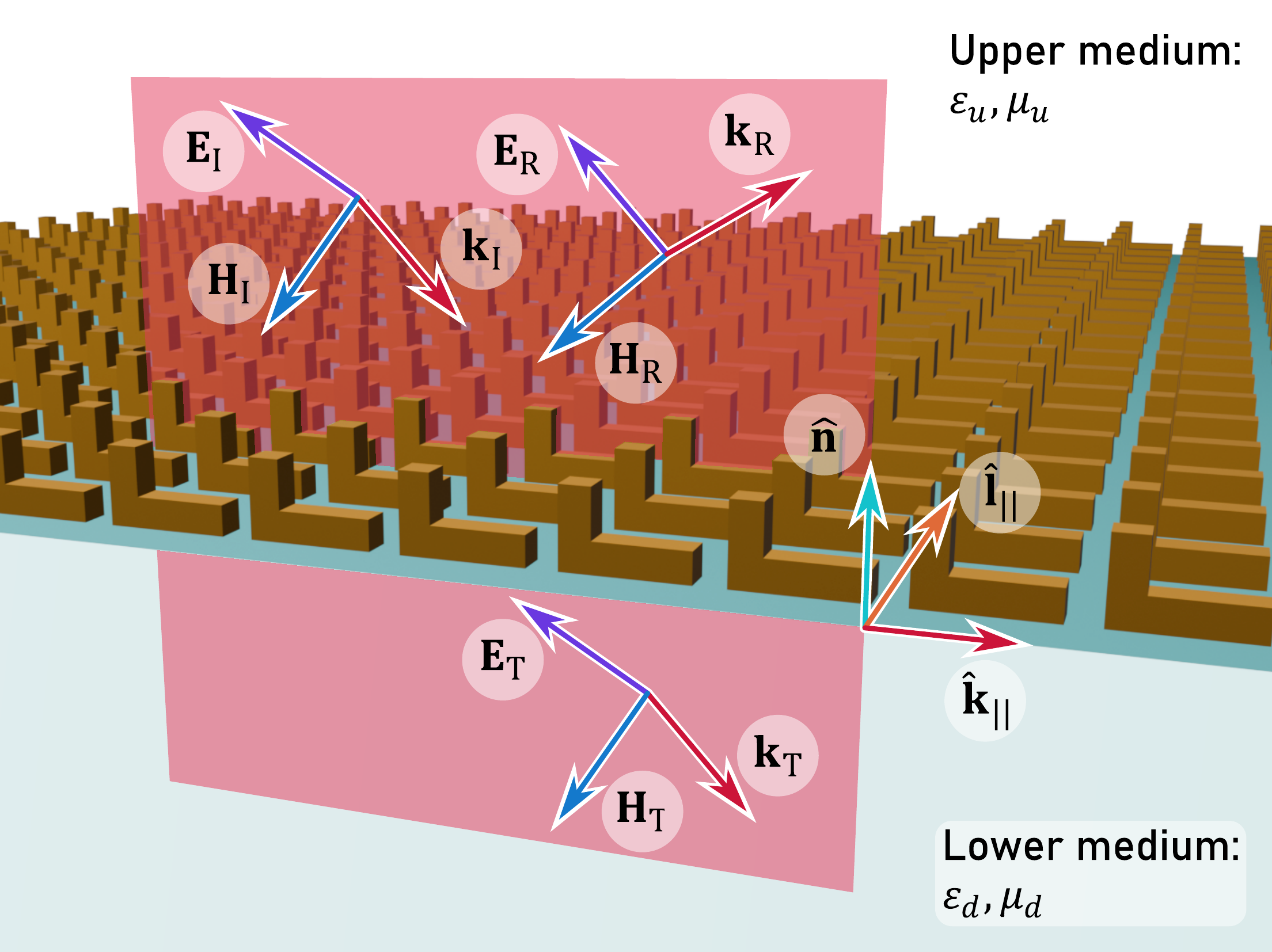}
\caption{Example of a generic scenario, where a metasurface is illuminated from above, producing transmitted and reflected waves of arbitrary polarization. In this scenario, the metasurface is surrounded by the arbitrary media above denoted as ``$u$'' and below as ``$d$'', where the red cut shows the propagation plane (normal to $\uvect{l}_\tg$). }\label{fig:schematic_fields}
\end{figure}

The main appeal of metasurfaces (and metamaterials in general) is its design philosophy, where the goal is to determine the unit-cell properties (geometry, materials) that produces the metasurface intended functionality. For such purpose, the metasurface is conceptualized as a sheet of negligible thickness and homogenized in terms of effective parameters, such as collective polarizabilities, surface impedances or susceptibilities~\cite{Simovski_Tretyakov_2020_metamaterials,Achouri2021_book}. Under this abstraction layer, the metasurface features are determined by solving the Generalized Sheet Transition Conditions (GSTCs) considering the illuminating source, the intended scattering and the surrounding media, as portrayed in Fig.~\ref{fig:schematic_fields}.

The main drawback of this analytical approach is the complexity of solving the GSTCs for each combination of incident and scattering waves (design) or the metasurface response for a given illumination (characterization). In particular for scattering analysis, the expressions for the metasurface transmission and reflection coefficients become increasingly more complex as more effective parameters are taken into account. Alternative approaches have been considered to address this issue. For instance, some recent research in metasurface design has been focused on the use of machine-learning for their characterization \cite{Liu_2018,Khoram_2023,Qiu_2019,Ueno_2024,Dong_2025}. However, this strategy is not viable for a general case as it requires plenty of numerical data (from either simulations or measurements) and computational resources for its training. On the other hand, a generalized analytical solution for the GSTCs would be more effective as it would require less computational resources. 

In that regards, a generalized solution for the GSTCs has been developed in Ref.~\cite{Albooyeh2016_general}, where the metasurface has been modeled in terms of collective polarizabilities; with its advantages and drawbacks. The main benefit of using collective polarizabilities is that it considers the behavior of the unit-cell as a single scatter and their interactions with neighboring unit-cells~\cite{Tretyakov2003_book,Simovski_Tretyakov_2020_metamaterials}. Unfortunately, while determining the polarizabilities for a single scatter is straight-forward in most cases, the interaction between particles is not, as some features can only being explained by the interaction of the metasurface as a whole. On the other hand, a general solution based on susceptibilities can be easier to use as its parameters model the metasurface as a single entity, while the complexity of the susceptibility tensors can be reduced by exploiting the symmetries of the unit-cell~\cite{Achouri2020_angular,Achouri2022_spatial}.

In this work, we present an analytical model that allows determining the scattering produced by a uniform metasurface homogenized in terms of effective susceptibilities. This works is divided into two sections. The first one presents the model as a generalized solution of the Generalized Sheet Transition Conditions for susceptibilities. While Sec.~\ref{sec:III} illustrates the method with different examples where the proposed model can be used to extract the equivalent susceptibilities for a given metasurface from its scattering coefficients, and to use such susceptibilities to analyze its theoretical behavior under different illumination conditions.

\section{Solution for the general case based on susceptibilities}\label{sec:II}
\subsection{Boundary conditions for arbitrary susceptibilities}\label{sec:II.1}

Consider the scenario of Fig.~\ref{fig:schematic_fields}, where a uniform and homogenizable sheet of negligible thickness with respect to the wavelength is placed between two media of arbitrary but isotropic characteristics. Under the illumination of a plane wave, the resulting scattering can be determined through the generalized sheet transition conditions (colloquially called boundary conditions in this work from now on)
\begin{subequations}
\label{eq:boundary_conditions}
\begin{equation}
\Delta\vect{E}_{\tg}= j\omega \uvect{n}\times \vect{M}_{\tg} - \nabla_{\tg} P_{n}/\varepsilon_0 ,
\end{equation}
\begin{equation}
\uvect{n}\times \Delta\vect{H}_{\tg} = j \omega \vect{P}_{\tg} + \nabla_{\tg}\times \left(\uvect{n} M_{n}/\mu_0\right),
\end{equation}
\end{subequations}
where $\Delta\vect{E}_{\tg}=\vect{E}_{\tg,u}-\vect{E}_{\tg,d}$ is the difference between the total tangential electric field at both sides of the sheet (the medium above the metasurface denoted as ``$u$'', and ``$d$'' for the medium below), $\Delta\vect{H}_{\tg}=\vect{H}_{\tg,u}-\vect{H}_{\tg,d}$ is the difference of the total tangential magnetic field, $\vect{P}_{\tg}$ ($\vect{M}_{\tg}$) is the tangential component of the equivalent surface electric (magnetic) polarization density, while  $P_{n}$ ($ M_{n}$) is the normal component of the surface electric (magnetic) polarization density with respect the vector normal to the sheet $\uvect{n}$~\cite{Albooyeh2016_general,Achouri2021_book,Simovski_Tretyakov_2020_metamaterials}.
The surface polarization densities can be modeled under different approaches ~\cite{Achouri2021_book,Simovski_Tretyakov_2020_metamaterials}, nevertheless, this work will consider the model based on susceptibilities
\begin{subequations}
\label{eq:susceptibilies}
\begin{equation}
\vect{P}=\varepsilon_0\dyad{\chi}\@{ee}\cdot \vect{E}\@{avg}+\dfrac{1}{c_0}\dyad{\chi}\@{em}\cdot \vect{H}\@{avg} ,
\end{equation}
\begin{equation}
\vect{M}=\mu_0\dyad{\chi}\@{mm}\cdot \vect{H}\@{avg}+\dfrac{1}{c_0}\dyad{\chi}\@{me}\cdot \vect{E}\@{avg},
\end{equation}
\end{subequations}
where $\dyad{\chi}$ is the susceptibility tensor for either electric (ee), magnetic (mm), electromagnetic (me) and magnetoelectric (em) coupling; and $\vect{E}\@{avg}$ ($\vect{H}\@{avg}$) is the averaged electric (magnetic) field at both sides of the sheet~\cite{Holloway2012_susceptibilities,Achouri2020_angular} defined as
\begin{subequations}
\label{eq:averaged_fields}
    \begin{equation}
        \vect{E}\@{avg}=\dfrac{1}{2}\sum_{F={\rm I,R,T}}\vect{E}_{\tg, F}+\dfrac{D_{n, F}}{\varepsilon_0}\uvect{n}
    \end{equation}
        \begin{equation}
        \vect{H}\@{avg}=\dfrac{1}{2}\sum_{F={\rm I,R,T}}\vect{H}_{\tg, F}+\dfrac{B_{n, F}}{\mu_0}\uvect{n},
    \end{equation}
\end{subequations}
where the subindex $F$ refers to the wave, either Incident (${\rm I}$), Transmitted (${\rm T}$), or Reflected (${\rm R}$); with tangential electric (magnetic) fields $\vect{E}_{\tg, F}$ ($\vect{H}_{\tg, F}$) and normal electric (magnetic) flux densities $D_{n, F}$ ($B_{n, F}$). 
The use of the normal components of the electric and magnetic flux densities in Eqs.~\eqref{eq:averaged_fields} comes from satisfying the boundary conditions related to the electric $\vect{D}$ and magnetic $\vect{B}$ flux densities ($E_n\rightarrow D_n/\varepsilon_0$ and $H_n\rightarrow B_n/\mu_0$)~\cite{Tiukuvaara2023_quadrupolar}.

\subsection{Coupling susceptibilities}\label{sec:II.2}

It can be noticed from Eqs.~\eqref{eq:susceptibilies} that the surface polarization densities can be linearly split in terms of the incident/scattering waves. In addition, due to Maxwell's equations in an isotropic medium, it is possible to define the normal and tangential components of the electric and magnetic field vectors in terms of the tangential part of the electric field~\cite{Albooyeh2016_general}. Therefore, the surface electric polarization density may be expressed as
\begin{subequations}
\label{eq:pdip_rho}
\begin{equation}
\vect{P}=\sum_{F={\rm I,R,T}} \vect{P}_{\tg,F} + P_{n,F}\uvect{n} ,
\end{equation}
\begin{equation}
\vect{P}_{\tg,F}=\dyad{V}_{\tg,F}\cdot\vect{E}_{\tg,F} ,
\end{equation}
\begin{equation}
P_{n,F}=\vect{V}_{n,F}\cdot\vect{E}_{\tg,F},
\end{equation}
\end{subequations}
where $\dyad{V}_{\tg,F}$ is the electro-electro-coupling susceptibility tangential tensor and $\vect{V}_{n,F}$ the electro-electro-coupling susceptibility normal vector. The role of $\dyad{V}_{\tg,F}$ and $\vect{V}_{n,F}$ is to characterize the contribution of the incident/scatter wave $F$ to the surface electric polarization in terms of the tangential electric field $\vect{E}_{\tg,F}$. Likewise, the surface magnetic polarization density can be written as
\begin{subequations}
\label{eq:mdip_nu}
\begin{equation}
\vect{M}=\sum_{F={\rm I,R,T}} \vect{M}_{\tg,F} + M_{n,F} \uvect{n} ,
\end{equation}
\begin{equation}
\vect{M}_{\tg,F}=\dyad{W}_{\tg,F}\cdot\vect{E}_{\tg,F} ,
\end{equation}
\begin{equation}
M_{n,F}=\vect{W}_{n,F}\cdot\vect{E}_{\tg,F},
\end{equation}
\end{subequations}
where $\dyad{W}_{\tg,F}$ ($\vect{W}_{n,F}$) is the electro-magnetic-coupling susceptibility that characterizes the tangential (normal) component of the surface magnetic polarization in terms of the tangential electric field.

The next part of the analysis will require to choose a particular coordinate axis system, nevertheless, the results can be implemented for any arbitrary system. As displayed in Fig.~\ref{fig:schematic_fields}, the first axis of this system is $\uvect{n}$, defined as normal to the sheet; while the second axis, $\uvect{k}_{\tg}$, is taken from the incident wave wavevector as
\begin{subequations}
\begin{equation}
\vect{k}_{F}=\vect{k}_{\tg}+k_{n,F} \uvect{n} ,
\end{equation}
\begin{equation}
\uvect{k}_{\tg}=\vect{k}_{\tg}/k_{\tg},\label{eq:ukt}
\end{equation}
\end{subequations}
where $k_{\tg}$ is the scalar value of the wavevector tangential component (equal for all waves due to Snell Law for specular scattering), and $k_{n,F}$ is the normal component of the wavevector, unique for each wave. The last axis $\uvect{l}_{\tg}$ is purposely defined as orthogonal to both $\uvect{n}$ and $\uvect{k}_{\tg}$
\begin{subequations}
\begin{equation}
\vect{l}_{\tg}=\uvect{n}\times\vect{k}_{\tg}\label{eq:lt}
\end{equation}
\begin{equation}
\uvect{l}_{\tg}=\vect{l}_{\tg}/k_{\tg}=\uvect{n}\times\uvect{k}_{\tg}.\label{eq:ult}
\end{equation}
\end{subequations}
Under this convention, both surface electric and magnetic polarization densities of Eqs.~\eqref{eq:pdip_rho}-\eqref{eq:mdip_nu} are split into their tangential and normal components. For such purpose, in a similar fashion as presented in Ref.~\cite{Albooyeh2016_general} for the effective polarizability tensors $\dyad{\hat{\alpha}}$, the susceptibilities tensors $\dyad{\chi}$ are expressed in the form
\begin{subequations}
\label{eq:susceptibility_decomposition}
\begin{equation}
\dyad{\chi}=
\left[\begin{array}{cc|c}
\chi^{kk} & \chi^{kl} & \chi^{kn}\\
\chi^{lk} & \chi^{ll} & \chi^{ln}\\
\hline
\chi^{nk} & \chi^{nl} & \chi^{nn}
\end{array}\right]
=
\left[\begin{array}{c|c}
\dyad{\chi}_{\tg} & \dyad{\chi}_{\tg n}\\
\hline
\dyad{\chi}_{n \tg} & \chi^{nn}
\end{array}\right] ,
\end{equation}
\begin{equation}
\dyad{\chi}_{\tg}=
\begin{bmatrix}
\chi^{kk} & \chi^{kl}\\
\chi^{lk} & \chi^{kk}
\end{bmatrix} ,
\end{equation}
\begin{equation}
\vect{\chi}_{\tg n} = \dyad{\chi}_{\tg n}\cdot\uvect{n} = \chi^{kn} \uvect{k}_{\tg} + \chi^{ln} \uvect{l}_{\tg} ,
\end{equation}
\begin{equation}
\vect{\chi}_{n \tg} = \uvect{n}\cdot\dyad{\chi}_{n \tg} = \chi^{nk} \uvect{k}_{\tg} + \chi^{nl} \uvect{l}_{\tg} .
\end{equation}
\end{subequations}
Therefore, the surface polarization densities read
\begin{subequations}
\label{eq:pdip_sus}
\begin{equation}
\vect{P}_{\tg,F}=\dfrac{1}{2}\Bigg[\varepsilon_0\dyad{\chi}_{{\rm ee},\tg}\cdot\vect{E}_{\tg,F}+\dfrac{1}{c_0}\dyad{\chi}_{{\rm em},\tg}\cdot\vect{H}_{\tg,F}+\varepsilon_0\vect{\chi}_{{\rm ee},\tg n}\dfrac{D_{n,F}}{\varepsilon_0}+\dfrac{1}{c_0}\vect{\chi}_{{\rm em},\tg n}\dfrac{B_{n,F}}{\mu_0}\Bigg] ,
\end{equation}
\begin{equation}
P_{n,F}=\dfrac{1}{2}\Bigg[\varepsilon_0\vect{\chi}_{{\rm ee},n\tg}\cdot\vect{E}_{\tg,F}+\dfrac{1}{c_0}\vect{\chi}_{{\rm em},n\tg}\cdot\vect{H}_{\tg,F} + \varepsilon_0{\chi\@{ee}^{nn}} \dfrac{D_{n,F}}{\varepsilon_0}+\dfrac{1}{c_0}{\chi\@{em}^{nn}}\dfrac{B_{n,F}}{\mu_0}\Bigg] ,
\end{equation}
\end{subequations}
\begin{subequations}
\label{eq:mpdip_sus}
\begin{equation}
\vect{M}_{\tg,F}=\dfrac{1}{2}\Bigg[\dfrac{1}{c_0}\dyad{\chi}_{{\rm me},\tg}\cdot\vect{E}_{\tg,F}+\mu_0\dyad{\chi}_{{\rm mm},\tg}\cdot\vect{H}_{\tg,F}+\dfrac{1}{c_0}\vect{\chi}_{{\rm me},\tg n}\dfrac{D_{n,F}}{\varepsilon_0}+\mu_0\vect{\chi}_{{\rm mm},\tg n}\dfrac{B_{n,F}}{\mu_0}\Bigg] ,
\end{equation}
\begin{equation}
M_{n,F}=\dfrac{1}{2}\Bigg[\dfrac{1}{c_0}\vect{\chi}_{{\rm me},n\tg}\cdot\vect{E}_{\tg,F}+\mu_0\vect{\chi}_{{\rm mm},n\tg}\cdot\vect{H}_{\tg,F}+\dfrac{1}{c_0}{\chi\@{me}^{nn}}\dfrac{D_{n,F}}{\varepsilon_0}+\mu_0{\chi\@{mm}^{nn}}\dfrac{B_{n,F}}{\mu_0}\Bigg] .
\end{equation}
\end{subequations}

The expressions for $D_{n}$, $B_{n}$ and $\vect{H}_\tg$ are obtained following Ref.~\cite{Albooyeh2016_general}, using the concept of wave impedance tensor $\dyad{Z}$. For convenience, this works uses the wave admittance vector $\dyad{Y}=\dyad{Z}^{-1}$ instead, which reads
\begin{subequations}
\begin{equation}
\dyad{Y}_{(u,d)}=\dfrac{1}{Z_{{\rm TM},(u,d)}} {\uvect{k}_{\tg}\uvect{k}_{\tg}} + \dfrac{1}{Z_{{\rm TE},(u,d)}} {\uvect{l}_{\tg}\uvect{l}_{\tg}} ,\label{eq:wave_admittance}
\end{equation}
\begin{equation}
Z_{{\rm TM},(u,d)}=\eta_{(u,d)} \sqrt{1-\left(k_{\tg}/k_{(u,d)}\right)^2} ,
\end{equation}
\begin{equation}
Z_{{\rm TE},(u,d)}=\dfrac{\eta_{(u,d)}}{\sqrt{1-\left(k_{\tg}/k_{(u,d)}\right)^2}},
\end{equation}
\end{subequations}
where the subindex $(u,d)$ corresponds to the medium in which the incident/scattered waves propagates, with characteristic impedance $\eta_{(u,d)}$ and wavenumber $k_{(u,d)}$. Each wave is expressed as a combination of a Transverse Electric (TE) and a Transverse Magnetic (TM) wave, with $Z\@{TE}$ and $Z\@{TM}$ representing the wave impedance for each kind of wave, respectively. The normal components of the electric and magnetic fluxes ($D_{n}$ and $B_{n}$), and the tangential component of the magnetic field $\vect{H}_\tg$ can be expressed in terms of the tangential electric field as
\begin{subequations}
\label{eq:fields_e_tg}
\begin{equation}
D_{n,F}/\varepsilon_0 =\varepsilon_{r,(u,d)}E_{n,F}=-\dfrac{{\rm sgn}\left(k_{n,F}\right)}{\omega \varepsilon_0 Z_{{\rm TM},(u,d)}} \vect{k}_{\tg}\cdot \vect{E}_{\tg,F} ,
\end{equation}
\begin{equation}
B_{n,F}/\mu_0 =\mu_{r,(u,d)}H_{n,F}= \dfrac{1}{\omega \mu_0} \vect{l}_{\tg}\cdot\vect{E}_{\tg,F} ,
\end{equation}
\begin{equation}
\vect{H}_{\tg,F}={\rm sgn}\left(k_{n,F}\right) \uvect{n}\times \dyad{Y}_{(u,d)} \cdot \vect{E}_{\tg,F},\label{eq:etg_to_htg}
\end{equation}
\end{subequations}
here, the signum function ``${\rm sgn}$'' links certain field components to the direction each wave propagates with respect to $\uvect{n}$. Using the expressions of Eqs.~\eqref{eq:fields_e_tg}, the coupling susceptibilities achieve the form
\begin{subequations}
\label{eq:rho_nu}
\begin{equation}
\dyad{V}_{\tg,F}=\dfrac{1}{2}\Bigg[\varepsilon_0\dyad{\chi}_{{\rm ee},\tg}-\dfrac{1}{c_0}\dyad{\chi}_{{\rm em},\tg}\cdot\left(\uvect{n}\times \dyad{Y}_{F}\right) +\dfrac{1}{\omega Z_{{\rm TM},F}}{\vect{\chi}_{{\rm ee},\tg n}\vect{k}_{\tg}}+\dfrac{1}{\omega \eta_0}{\vect{\chi}_{{\rm em},\tg n}\vect{l}_{\tg}}\Bigg] ,
\end{equation}
\begin{equation}
\vect{V}_{n,F}=\dfrac{1}{2}\Bigg[\varepsilon_0\vect{\chi}_{{\rm ee},n\tg}-\dfrac{1}{c_0}\vect{\chi}_{{\rm em},n\tg}\cdot\left(\uvect{n}\times \dyad{Y}_{F}\right) +\dfrac{{\chi\@{ee}^{nn}}}{\omega Z_{{\rm TM},F}} \vect{k}_{\tg}+\dfrac{{\chi\@{em}^{nn}}}{\omega \eta_0} \vect{l}_{\tg}\Bigg] ,
\end{equation}
\begin{equation}
\dyad{W}_{\tg,F}=\dfrac{1}{2}\Bigg[\dfrac{1}{c_0}\dyad{\chi}_{{\rm me},\tg}-\mu_0\dyad{\chi}_{{\rm mm},\tg}\cdot\left(\uvect{n}\times \dyad{Y}_{F}\right)+\dfrac{\eta_0}{\omega Z_{{\rm TM},F}}{\vect{\chi}_{{\rm me},\tg n}\vect{k}_{\tg}}+\dfrac{1}{\omega}{\vect{\chi}_{{\rm mm},\tg n}\vect{l}_{\tg}}\Bigg] ,
\end{equation}
\begin{equation}
\vect{W}_{n,F}=\dfrac{1}{2}\Bigg[\dfrac{1}{c_0}\vect{\chi}_{{\rm me},n\tg}-\mu_0\vect{\chi}_{{\rm mm},n\tg}\cdot\left(\uvect{n}\times \dyad{Y}_{F}\right)+\dfrac{\eta_0{\chi\@{me}^{nn}}}{\omega Z_{{\rm TM},F}} \vect{k}_{\tg}+\dfrac{{\chi\@{mm}^{nn}}}{\omega} \vect{l}_{\tg}\Bigg],
\end{equation}
\end{subequations}
where the terms $\dyad{Y}_{F}$ and $Z_{{\rm TM},F}$ depends both on the wave (incident, reflected, transmitted) and the direction of the incident wave (going from the upper medium to the lower one, in $-\uvect{n}$ direction, or vice versa in $+\uvect{n}$ direction). Each variable reads
\begin{subequations}
\begin{equation}
\dyad{Y}_{F}=\left\{ \begin{array}{c l l}
\pm \dyad{Y}_{(u,d)} & {\rm for} & F={\rm I}\\
\mp \dyad{Y}_{(u,d)} & {\rm for} & F={\rm R}\\
\pm \dyad{Y}_{(d,u)} & {\rm for} & F={\rm T}
\end{array}\right. ,
\end{equation}
\begin{equation}
Z_{{\rm TM},F}=\left\{ \begin{array}{c l l}
\pm Z_{{\rm TM},(u,d)} & {\rm for} & F={\rm I}\\
\mp Z_{{\rm TM},(u,d)} & {\rm for} & F={\rm R}\\
\pm Z_{{\rm TM},(d,u)} & {\rm for} & F={\rm T}
\end{array}\right. ,
\end{equation}
\end{subequations}
where the first subindex and top sign correspond to illumination from the upper medium ($-\uvect{n}$ propagation), while the second subindex and bottom sign characterized illumination coming from the lower medium.

\subsection{Scattering coefficients for arbitritrary susceptibilities}\label{sec:II.3}
By considering the polarization tensors and vectors of Eqs.~\eqref{eq:rho_nu}, the boundary conditions of Eqs.~\eqref{eq:boundary_conditions} take the form
\begin{subequations}
\label{eq:boundary_conditions_rho_nu}
\begin{equation}
\Delta\vect{E}_{\tg} = \sum_{F={\rm I,R,T}}\left[j\omega \uvect{n}\times \dyad{W}_{\tg,F} +j \dfrac{1}{\varepsilon_0} {\vect{k}_{\tg} \vect{V}_{n,F}}\right]\cdot\vect{E}_{\tg,F} ,\label{eq:boundary_conditions_rho_nu_e}
\end{equation}
\begin{equation}
\uvect{n}\times \Delta\vect{H}_{\tg} = \sum_{F={\rm I,R,T}} \left[j \omega \dyad{V}_{\tg,F} +j \dfrac{1}{\mu_0} {\vect{l}_{\tg}\vect{W}_{n,F}}\right]\cdot\vect{E}_{\tg,F},\label{eq:boundary_conditions_rho_nu_h}
\end{equation}
\end{subequations}
where the operator $\nabla_{\tg}\rightarrow -j \vect{k}_{\tg}$ [using the time harmonic convention $\exp (+j\omega t)$]. The equation system presented above has two unknowns $\vect{E}\@{\tg,R}$ and $\vect{E}\@{\tg,T}$ [the tangential magnetic fields can be obtained using Eq.~\eqref{eq:etg_to_htg}]. For such purpose, the boundary conditions of Eqs.~\eqref{eq:boundary_conditions_rho_nu} can be rearranged, such that they take the form 
\begin{subequations}
\label{eq:boundary_conditions_with_bees}
\begin{equation}
\dyad{B}\@{E,I}\cdot \vect{E}\@{\vert\vert,I}+\dyad{B}\@{E,R}\cdot \vect{E}\@{\vert\vert,R}+\dyad{B}\@{E,T}\cdot \vect{E}\@{\vert\vert,T}=0 , 
\end{equation}
\begin{equation}
\dyad{B}\@{H,I}\cdot \vect{E}\@{\vert\vert,I}+\dyad{B}\@{H,R}\cdot \vect{E}\@{\vert\vert,R}+\dyad{B}\@{H,T}\cdot \vect{E}\@{\vert\vert,T}=0.
\end{equation}
\end{subequations}
The auxiliary tensors $\dyad{B}\@{E}$ are obtained from rearranging the tangential electric fields on the boundary conditions of Eq.~\eqref{eq:boundary_conditions_rho_nu_e}
\begin{subequations}
\begin{equation}
\dyad{B}\@{E,I}=\pm\dyad{I}_{\tg}-j\left[\omega \uvect{n}\times \dyad{W}_{\tg,\rm I} +\dfrac{1}{\varepsilon_0} {\vect{k}_{\tg} \vect{V}_{n,\rm I}}\right] ,
\end{equation}
\begin{equation}
\dyad{B}\@{E,R}=\pm\dyad{I}_{\tg}-j\left[\omega \uvect{n}\times \dyad{W}_{\tg,\rm R} +\dfrac{1}{\varepsilon_0} {\vect{k}_{\tg} \vect{V}_{n,\rm R}}\right] ,
\end{equation}
\begin{equation}
\dyad{B}\@{E,T}=\mp\dyad{I}_{\tg}-j\left[\omega \uvect{n}\times \dyad{W}_{\tg,\rm T} +\dfrac{1}{\varepsilon_0} {\vect{k}_{\tg} \vect{V}_{n,\rm T}}\right],
\end{equation}
\end{subequations}
where $\dyad{I}_{\tg}=\dyad{I}-\uvect{n}\uvect{n}$ is the unitary two-dimensional tensor in the plane defined by $\uvect{n}$. Similarly, $\dyad{B}\@{H}$ are obtained from Eq.~\eqref{eq:boundary_conditions_rho_nu_h}
\begin{subequations}
\begin{equation}
\dyad{B}\@{H,I}=\dyad{Y}_{(u,d)}-j\left[\omega \dyad{V}_{\tg,\rm I} +\dfrac{1}{\mu_0} {\vect{l}_{\tg}\vect{W}_{n,\rm I}}\right] ,
\end{equation}
\begin{equation}
\dyad{B}\@{H,R}=-\dyad{Y}_{(u,d)}-j\left[\omega \dyad{V}_{\tg,\rm R} +\dfrac{1}{\mu_0} {\vect{l}_{\tg}\vect{W}_{n,\rm R}}\right] ,
\end{equation}
\begin{equation}
\dyad{B}\@{H,T}=-\dyad{Y}_{(d,u)}-j\left[\omega \dyad{V}_{\tg,\rm T} +\dfrac{1}{\mu_0} {\vect{l}_{\tg}\vect{W}_{n,\rm T}}\right].
\end{equation}
\end{subequations}
The solution of Eqs.~\eqref{eq:boundary_conditions_with_bees} consist of two tangential scattering tensors $\dyad{T}_{\tg}$ and $\dyad{R}_{\tg}$ that satisfies the relations 
\begin{subequations}
\label{eq:scatt_tang_tensor}
\begin{equation}
\vect{E}_{\tg,\rm T}=\dyad{T}_{\tg}\cdot \vect{E}_{\tg,\rm I} ,
\end{equation}
\begin{equation}
\vect{E}_{\tg,\rm R}=\dyad{R}_{\tg}\cdot \vect{E}_{\tg,\rm I}.
\end{equation}
\end{subequations}
Solving the expression of Eqs.~\eqref{eq:boundary_conditions_with_bees} is comparable to figuring out a two-equation system with two unknowns ($\vect{E}_{\tg,\rm T}$ and $\vect{E}_{\tg,\rm R}$). Because of that, the expressions for $\dyad{T}_{\tg}$ and $\dyad{R}_{\tg}$ read
\begin{subequations}
\label{eq:tau_tensor}
\begin{equation}
\dyad{T}_{\tg}=-\dyad{A}\@{T,1}^{-1}\cdot\dyad{A}\@{T,2} ,
\end{equation}
\begin{equation}
\dyad{A}\@{T,1}=\dyad{B}\@{E,T}-\dyad{B}\@{E,R}\cdot\dyad{B}\@{H,R}^{-1}\cdot\dyad{B}\@{H,T} ,
\end{equation}
\begin{equation}
\dyad{A}\@{T,2}=\dyad{B}\@{E,I}-\dyad{B}\@{E,R}\cdot\dyad{B}\@{H,R}^{-1}\cdot\dyad{B}\@{H,I} ,
\end{equation}
\end{subequations}
and
\begin{subequations}
\label{eq:gamma_tensor}
\begin{equation}
\dyad{R}_{\tg}=-\dyad{A}\@{R,1}^{-1}\cdot\dyad{A}\@{R,2} ,
\end{equation}
\begin{equation}
\dyad{A}\@{R,1}=\dyad{B}\@{E,R}-\dyad{B}\@{E,T}\cdot\dyad{B}\@{H,T}^{-1}\cdot\dyad{B}\@{H,R} ,
\end{equation}
\begin{equation}
\dyad{A}\@{R,2}=\dyad{B}\@{E,I}-\dyad{B}\@{E,T}\cdot\dyad{B}\@{H,T}^{-1}\cdot\dyad{B}\@{H,I}.
\end{equation}
\end{subequations}
%

%
%

The next step consists in converting the scattering tensors for tangential electric fields $\dyad{T}_{\tg}$, $\dyad{R}_{\tg}$ into tensors that relates normalized outgoing and incident waves $\dyad{T}$, $\dyad{R}$~\cite{Cuesta_invisible_cavities_2018}. For such purpose, the wave scattering tensors for linearly-polarized waves (TM and TE) are written as
\begin{subequations}
\label{eq:scatt_coeff}
\begin{equation}
\dyad{T}= \dyad{Y}_{(d,u)}^{1/2}\cdot\dyad{T}_{\vert\vert}\cdot \dyad{Z}_{(u,d)}^{1/2} ,
\end{equation}
\begin{equation}
\dyad{R}= \dyad{Y}_{(u,d)}^{1/2}\cdot\dyad{R}_{\vert\vert}\cdot \dyad{Z}_{(u,d)}^{1/2} ,
\end{equation}
\end{subequations}
with
\begin{subequations}
\label{eq:norm_imp_tensors}
\begin{equation}
\dyad{Z}_{(u,d)}^{1/2} \triangleq \sqrt{Z_{{\rm TM},(u,d)}} \uvect{k}_{\vert\vert}\uvect{k}_{\vert\vert} + \sqrt{Z_{{\rm TE},(u,d)}} \uvect{l}_{\vert\vert}\uvect{l}_{\vert\vert} ,
\end{equation}
\begin{equation}
\dyad{Y}_{(u,d)}^{1/2} \triangleq \sqrt{Y_{{\rm TM},(u,d)}} \uvect{k}_{\vert\vert}\uvect{k}_{\vert\vert} + \sqrt{Y_{{\rm TE},(u,d)}} \uvect{l}_{\vert\vert}\uvect{l}_{\vert\vert}.
\end{equation}
\end{subequations}
The purpose of Eqs.~\eqref{eq:scatt_coeff} is to normalize the tangential scattering coefficients with the respect to the wave impedance. Because of that, Eqs.~\eqref{eq:norm_imp_tensors} consider the square root of the wave impedance for each linear polarization, while applying the wave impedance corresponding to the scattered and incident polarizations.
The last procedure consists of extracting the individual scattering coefficients from the scattering tensors. This is done by using the inner (dot) product between the scattering tensor and the unitary vector co-linear to the tangential electric field ($\uvect{q}\@{TM}=\uvect{k}_{\tg}$ for TM polarization and $\uvect{q}\@{TE}=-\uvect{l}_{\tg}$ for TE) for the corresponding incident and scattering wave, with the forms
\begin{subequations}
\label{eq:scatt_coeff_scalar}
\begin{align}
T\@{out,in}&= \uvect{q}\@{out}\cdot \dyad{T}\cdot\uvect{q}\@{in}, \\
R\@{out,in}&= \uvect{q}\@{out}\cdot \dyad{R}\cdot\uvect{q}\@{in},
\end{align}
\end{subequations}
where the subindex ${\rm (out,in)}$ indicates the polarization of the incident ${\rm (in)}$ and scattered ${\rm (out)}$ waves.
%
%
%
%
\subsection{Scattering coefficients for circular polarization}\label{sec:II.4}

The results obtained in Eqs.~\eqref{eq:scatt_coeff_scalar} for linear polarization (LP) can be extended for circularly-polarized (CP) waves through the Jones vector~\cite{Azzam_Bashara_1977}. For that purpose, we can write transmission and reflection tensors as matrices, which in the case of linear-polarization read
\begin{subequations}
\label{eq:lin_coeff_matrix}
    \begin{equation}
        \left[{T}\@{LP}\right]=
        \begin{bmatrix}
        T\@{TM,TM} & T\@{TM,TE}\\
        T\@{TE,TM} & T\@{TE,TE}
        \end{bmatrix} ,
    \end{equation}
\begin{equation}
        \left[{R}\@{LP}\right]=
        \begin{bmatrix}
        R\@{TM,TM} & R\@{TM,TE}\\
        R\@{TE,TM} & R\@{TE,TE}
        \end{bmatrix}.
\end{equation}
\end{subequations}
Therefore, the scattering matrices for the circular-polarized cases are expressed as
\begin{subequations}
\label{eq:circ_coeff_matrix}
    \begin{equation}
        \left[T\@{CP}\right]=
        \begin{bmatrix}
        T\@{RCP,RCP} & T\@{RCP,LCP}\\
        T\@{LCP,RCP} & T\@{LCP,LCP}
        \end{bmatrix}
        = [U]^{-1}\cdot
        \begin{bmatrix}
        T\@{TM,TM} & T\@{TM,TE}\\
        T\@{TE,TM} & T\@{TE,TE}
        \end{bmatrix}\cdot
        [U] ,
    \end{equation}     
\begin{equation}
        \left[R\@{CP}\right]=
        \begin{bmatrix}
        R\@{RCP,RCP} & R\@{RCP,LCP}\\
        R\@{LCP,RCP} & R\@{LCP,LCP}
        \end{bmatrix}
        = [U]^{T}\cdot
        \begin{bmatrix}
        R\@{TM,TM} & R\@{TM,TE}\\
        R\@{TE,TM} & R\@{TE,TE}
        \end{bmatrix}\cdot
        [U],     
\end{equation}
\end{subequations}
where RCP and LCP stand for right-hand and left-hand circular polarizations, respectively; and $[U]$ (with $[U]^{T}$ as its transpose) is the transformation matrix, written as
\begin{equation}
    \left[U\right]=\dfrac{1}{\sqrt{2}}
        \begin{bmatrix}
        j & -j\\
        1 & 1
        \end{bmatrix}.
\end{equation}

By extension, circular-polarized waves produced by linearly-polarized waves (and vice-versa) can be determined by applying the transformation matrix only to one side of Eqs.~\eqref{eq:circ_coeff_matrix}. In the aforementioned case (circular-polarized scattering from linearly-polarized excitation), the scattering coefficients read
\begin{subequations}
\label{eq:circ_lin_coeff_matrix}
    \begin{equation}
       \begin{bmatrix}
        T\@{RCP,TM} & T\@{RCP,TE}\\
        T\@{LCP,TM} & T\@{LCP,TE}
        \end{bmatrix}
        = [U]^{-1}\cdot
        \begin{bmatrix}
        T\@{TM,TM} & T\@{TM,TE}\\
        T\@{TE,TM} & T\@{TE,TE}
        \end{bmatrix} ,
    \end{equation}
    \begin{equation}        
        \begin{bmatrix}
        R\@{RCP,TM} & R\@{RCP,TE}\\
        R\@{LCP,TM} & R\@{LCP,TE}
        \end{bmatrix}
        = [U]^{T}\cdot
        \begin{bmatrix}
        R\@{TM,TM} & R\@{TM,TE}\\
        R\@{TE,TM} & R\@{TE,TE}
        \end{bmatrix} .
    \end{equation}
\end{subequations}
The complementary case (linearly-polarized scattering from circular-polarized excitation) is obtained by keeping the right-side transformation matrix of Eqs.~\eqref{eq:circ_coeff_matrix} instead.

\section{Examples}\label{sec:III}

In this section, we have selected three metasurfaces with unique properties \cite{Tiukuvaara2022,Plum2009,Baena2017}. For each metasurface, the model developed in Sec.~\ref{sec:II} is applied to determine the effective susceptibilities required to emulate the metasurface response. Once each metasurface parameters are extracted, it is possible to predict their behavior for scenarios not covered during simulations nor measurements.

\subsection{Transmissive Angular-Asymmetric Metasurface}\label{sec:III_L_shape}

\begin{figure}[h]
\centering
\includegraphics[width=0.6\linewidth]{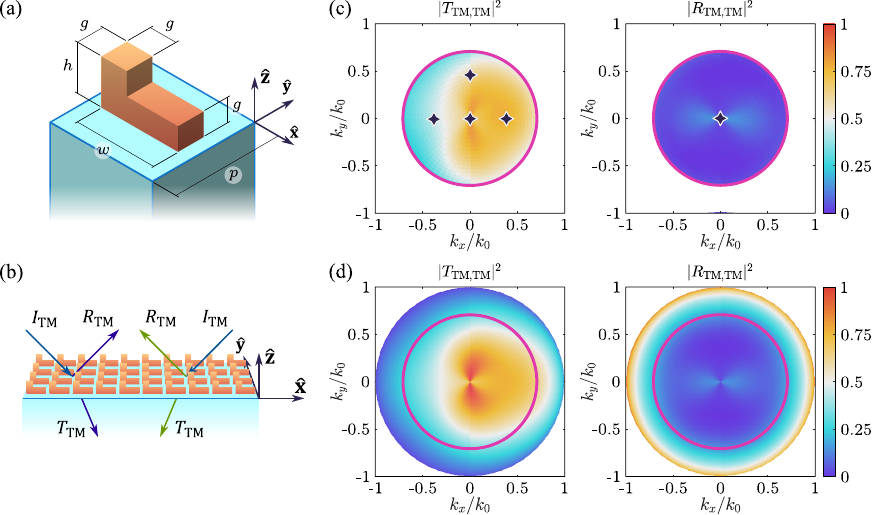}
\caption{(a) Schematic of the asymmetrically-transmissive metasurface unit-cell. (b) The analyzed metasurface has an asymmetric transmission along the $x$-axis when illuminated with TM-polarized waves. (c) Simulation results and sampling points (denoted as dark stars). The magenta line show the upper bound for the simulation results at $\theta=45^{\circ}$. (d) Extrapolated results based on the effective susceptibilities calculated from the sampling points. }\label{fig:l_shape_results}
\end{figure}

The first example consist in a metasurface which has an asymmetric response for co-polarized TM transmission for illumination in the $xz$ propagation plane, as represented in Fig.~\ref{fig:l_shape_results}(a)-(b) \cite{Tiukuvaara2022}. This metasurface was chosen for demonstration due to its tridimensional asymmetric unit-cell geometry (with normal-to-the-plane surface polarization densities) and the asymmetric media before and after the metasurface (which increases the analytical complexity of the system). Using the information and unit-cell dimensions provided in Ref.~\cite{Tiukuvaara2022} ($p=250$~[nm], $w=200$~[nm], $h=130$~[nm], $g=50$~[nm]), operational wavelength $\lambda_0=655$~[nm], and symmetry-allowed susceptibilities ($\chi\@{ee}^{xx}$, $\chi\@{ee}^{zz}$, $\chi\@{ee}^{xz}$, $\chi\@{mm}^{yy}$ and $\chi\@{em}^{xy}$) found by applying the methodology in~\cite{Achouri2022_spatial}. Using the analytical model in Sec.~\ref{sec:II} with the given set of susceptibilities, and adding the required susceptibilities to ensure the reciprocity of the metasurface, the co-polarized transmission coefficient for TM-illumination alongside the $xz$ propagation plane reads
\begin{subequations}
    \begin{equation}
        T\@{TM,TM}  =-\frac{2 n_d \sqrt{k_{z,u} k_{z,d}} }{D\@{TM,TM}}\left[k_0^2 \left({\chi\@{ee}^{xx}} {\chi\@{mm}^{yy}}+{\chi\@{em}^{xy}}^2\right)+k_x^2 \left({\chi\@{ee}^{xx}} {\chi\@{ee}^{zz}}-{\chi\@{ee}^{xz}}^2\right)+4 j k_x {\chi\@{ee}^{xz}}+4\right] ,
    \end{equation}
    \begin{equation}
    \begin{split}
        D\@{TM,TM}  = & k_{z,d} \left[k_0^2 {\chi\@{ee}^{xx}} {\chi\@{mm}^{yy}}+(k_0 {\chi\@{em}^{xy}}+2 j)^2+k_x^2 \left({\chi\@{ee}^{xx}} {\chi\@{ee}^{zz}}-{\chi\@{ee}^{xz}}^2\right)-4 j k_{z,u} {\chi\@{ee}^{xx}}\right]\\
        &+n_d^2 \left[k_{z,u} \left(k_0^2 {\chi\@{ee}^{xx}} {\chi\@{mm}^{yy}}+(k_0 {\chi\@{em}^{xy}}-2 j)^2+k_x^2 \left({\chi\@{ee}^{xx}} {\chi\@{ee}^{zz}}-{\chi\@{ee}^{xz}}^2\right)\right)-4 j \left(k_0^2 {\chi\@{mm}^{yy}}+k_x^2 {\chi\@{ee}^{zz}}\right)\right] ,
    \end{split}
    \end{equation}
\end{subequations}
where $n_d$ is the substrate refraction index, and $k_{z,(u,d)}=\sqrt{k_{(u,d)}^2-k_{\tg}^2}$. The metasurface was simulated in CST considering a unit-cell made of gold  \cite{Johnson_Christy_1972}, over a silicon dioxide substrate ($\varepsilon_r = 2.1316$, $n_d=\sqrt{\varepsilon_r}$), which numerical results for co-polarized TM transmission are displayed in Fig.~\ref{fig:l_shape_results}(c) (complete simulation results are presented in the Supporting Information). 

 \begin{table}[h]
\centering
\caption{Sampling scenarios and normalized effective susceptibilities for the L-shaped metasurface of Fig.~\ref{fig:l_shape_results}. }
\begin{tabular}{ c  c  c   c }
\hline
& Parameter & $\theta$~[deg] & $\phi$~[deg] \\ [0.5ex] 
\hline
1& $T\@{TM,TM}$ & 0 & 0 \\
2& $R\@{TM,TM}$ & 0 & 0 \\
3& $T\@{TM,TM}$ & 22.5 & 0 \\
4& $T\@{TM,TM}$ & 22.5 & 180 \\
5& $T\@{TM,TM}$ & 27.5 & 90 \\
\hline
\end{tabular}
\quad
\begin{tabular}{ c c }
\hline
& $\tilde{\chi}$ \\ [0.5ex] 
\hline
$\chi\@{ee}^{xx}$ & $-0.6472 + 0.9253j$ \\
$\chi\@{ee}^{zz}$ & $0.8301 - 1.946j$ \\
$\chi\@{ee}^{xz}$ & $0.1761 - 0.3066j$ \\
$\chi\@{mm}^{yy}$ & $0.2003 - 0.4127j$ \\
$\chi\@{em}^{xy}$ & $0.1454 - 0.7635j$ \\
\hline
\end{tabular}
\label{tab:L_shape_data}
\end{table}

\begin{figure}[h]
\centering
\includegraphics[width=0.45\linewidth]{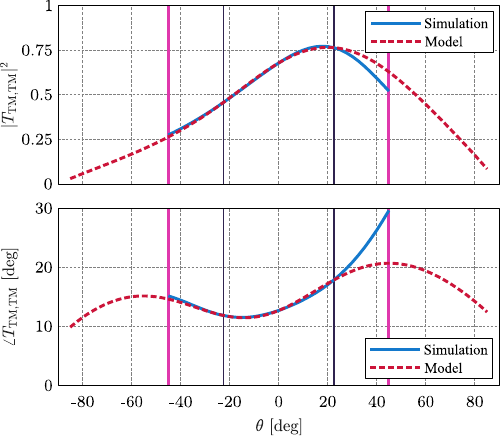}
\caption{Comparison between the simulated data and the predicted scattering in terms of $T\@{TM,TM}$ for illumination in the $xz$ plane for the metasurface of Fig.~\ref{fig:l_shape_results}. The magenta lines show the limits of the simulated data, while the dark lines denote the sampling poits with $\theta=\pm 22.5^{\circ}$.}\label{fig:l_shape_results_pr2}
\end{figure}

The equivalent susceptibilities values can be found by illuminating the metasurface from different directions, and for this example, the sampling cases are summarized in Tab.~\ref{tab:L_shape_data} and represented in Fig.~\ref{fig:l_shape_results}(c). The total number of sampling cases are equal to the number of unknown susceptibilities (five for the current unit-cell), and they were selected considering the intended metasurface features. At first, normal incidence case was considered as its scattering coefficients (transmission and reflection) do not depend on the out-of-plane susceptibilities ($\chi\@{ee}^{zz}$ and $\chi\@{ee}^{xz}$ for this metasurface), helping to characterize the in-plane susceptibilities. On the other hand, the sampling points 3 and 4 of Tab.~\ref{tab:L_shape_data} ($\theta=22.5^{\circ}$, $\phi=0^{\circ},180^{\circ}$) helps to characterize the asymmetry in $T\@{TM,TM}$, as portrayed in Fig.~\ref{fig:l_shape_results}(c). The sampling point at $\theta=22.5^{\circ},\phi=0^{\circ}$ was chosen since it is the closest to the $\vert T\@{TM,TM}\vert$ peak along the $x$ axis, as shown in Fig.~\ref{fig:l_shape_results_pr2}. The remaining sampling point at $\theta= 27.5^{\circ}$,~$\phi=90^{\circ}$ was chosen to characterize the metasurface for illumination in the $yz$ plane. 

Once the sampling points are established, the effective susceptibilities were calculated using \mbox{MATLAB} \verb+vpasolve+ function, using the analytical functions found from implementing the model of Sec.~\ref{sec:II} into Mathematica for illumination in both $xz$ and $yz$ planes. The effective susceptibilities for this example are summarized in Tab.~\ref{tab:L_shape_data} (with normalized values $\tilde{\chi}=k_0\chi$, being $k_0$ the free-space operational wavenumber), and results for transmission and reflection for TM illumination displayed in Fig.~\ref{fig:l_shape_results}(d) (a code example of the model implementation in MATLAB \cite{Cuesta2025-bp} and complementary scattering coefficient plots can be found in the Supporting Information).

By comparing Figs.~\ref{fig:l_shape_results}(c) and (d), it can be observed that the effective susceptibilities of Tab.~\ref{tab:L_shape_data} can replicate the metasurface behavior for $T\@{TM,TM}$ and $R\@{TM,TM}$ inside the simulated area, denoted by the magenta line at $\theta=45^{\circ}$. For illumination in the $xz$ propagation plane, as shown in Fig.~\ref{fig:l_shape_results_pr2}, it can be seen that the predicted values follow nearly perfectly the simulated data from $\theta=-45^{\circ}$ (the simulation lower bound) to $\theta=22.5^{\circ}$ (near $\vert T\@{TM,TM} \vert$ peak). The divergence between simulated data and  the predicted model for $\theta>22.5^{\circ}$, and the high values of $R\@{TM,TM}$ of Fig.~\ref{fig:l_shape_results}(d) for large $\theta$ values can be explained due to the excitation of higher-order multipoles, which are not considered in the boundary conditions of Eqs.~\eqref{eq:boundary_conditions}.

\subsection{Asymmetric Achiral Split-Ring with Optical Activity}\label{sec:III_split_ring}

\begin{figure}[h]
\centering
\includegraphics[width=1\linewidth]{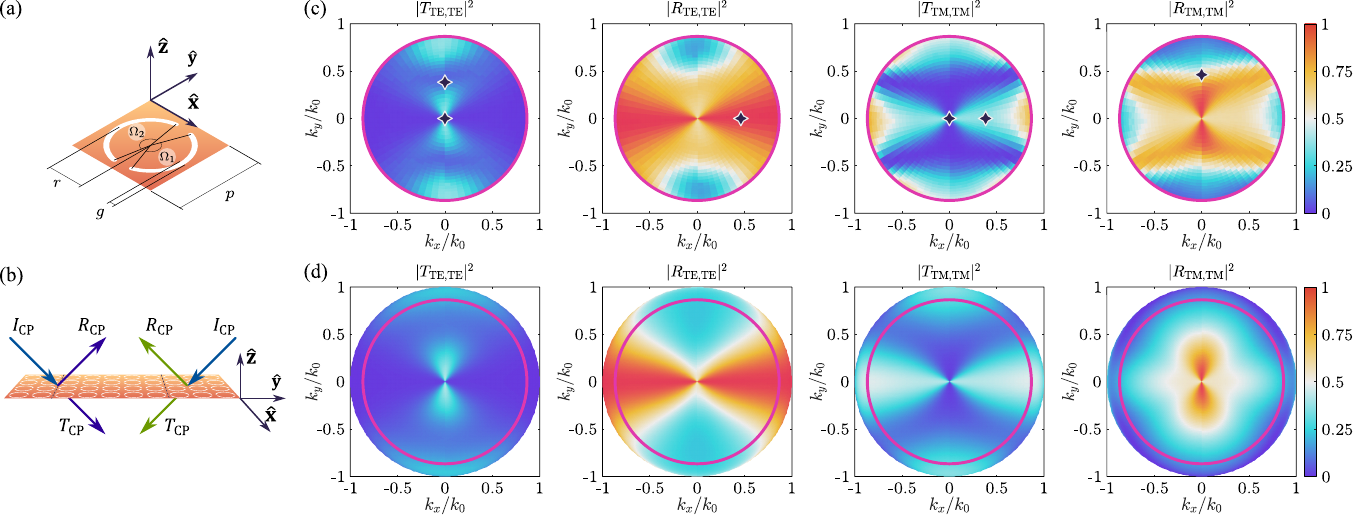}
\caption{(a) Schematic of the unit-cell corresponding with the metasurface with extrinsic chirality. Please note that the unit-cell is symmetric in the $y$ direction. (b) The metasurface behaves differently depending on the handiness of the circularly-polarized source when it is illuminated in the $yz$ plane. (c) Simulation results and sampling points (denoted as dark stars) for linearly-polarized illumination. The magenta line show the upper bound for the simulation results at $\theta=60^{\circ}$. (d) Predicted results based on the effective susceptibilities calculated from the sampling points.  }\label{fig:babinet_split_ring_results}
\end{figure}

The next example in this work, as presented in Fig.~\ref{fig:babinet_split_ring_results}(a)-(b), was selected due to its extrinsic chirality under circular polarization. Particularly, while the unit cell of this metasurface is symmetric along the $y$ axis, it will behave differently depending from which side in the $yz$ propagation plane when illuminated with a circularly-polarized wave~\cite{Plum2009,Achouri2022_spatial}. The unit-cell presented in Fig.~\ref{fig:babinet_split_ring_results}(a) follows the dimensions of Ref.~\cite{Plum2009} of an aluminum unit-cell with $p=15$~[mm], $r=6$~[mm], $g=1$~[mm], $\Omega_1=140^{\circ}$ and $\Omega_2=160^{\circ}$ for an operational frequency $f_0=8.49$~[GHz]. Since the unit-cell has symmetries $C_{2,x}$, $\sigma_y$ and $\sigma_z$; its equivalent susceptibility tensors can be reduced to~\cite{Achouri2022_spatial}
%
%
\begin{equation}
        \dyad{\chi}\@{ee,mm}=
        \begin{bmatrix}
        \chi^{xx} & 0 & 0\\
        0 & \chi^{yy} & 0\\
        0 & 0 & \chi^{zz}
        \end{bmatrix} , \quad   \dyad{\chi}\@{em,me}=
        \begin{bmatrix}
        0 & 0 & 0\\
        0 & 0 & \chi^{yz}\\
        0 & \chi^{zy} & 0
        \end{bmatrix}. \label{eq:split_ring_susceptibilities}
\end{equation}
Using a similar procedure to the one in Sec.~\ref{sec:III_L_shape}, the in-plane susceptibilities were characterized using normal-illumination, considering transmission and reflection for both TE and TM illuminations. From these results, it was found out that the susceptibilities $\chi\@{mm}^{xx}$ and $\chi\@{mm}^{yy}$ were negligible compared to the other in-plane susceptibilities. By using the proposed general solution, it can be found that the co-polarized transmission for circular-polarized waves in the $yz$ propagation plane is an odd function with respect $k_y$:
\begin{equation}
\begin{split}
    T\@{CP,CP} =& \frac{k_z }{k_z {\chi\@{ee}^{yy}}-2 j}\Bigg[\frac{-\left({k_y}^2 \left({\chi\@{em}^{zy}}^2+{\chi\@{ee}^{yy}} \left({\chi\@{ee}^{zz}}+j k_z {\chi\@{em}^{zy}}^2\right)\right)\right)\pm 2 {k_y} {\chi\@{em}^{zy}} (2+j k_z {\chi\@{ee}^{yy}})-4}{2 {k_y}^2 {\chi\@{ee}^{zz}}+j k_z \left({k_y}^2 {\chi\@{em}^{zy}}^2-4\right)}\\
    &-\frac{j (k_z {\chi\@{ee}^{yy}}+j ({\pm k_y} {\chi\@{em}^{yz}}-2))^2}{k_z \left({k_y}^2 \left({\chi\@{ee}^{yy}} ({\chi\@{ee}^{xx}}+{\chi\@{mm}^{zz}})+{\chi\@{em}^{yz}}^2\right)-4\right)-2 j {k_y}^2 ({\chi\@{ee}^{xx}}+{\chi\@{mm}^{zz}})+k_z^3 {\chi\@{ee}^{xx}} {\chi\@{ee}^{yy}}-2 j k_z^2 ({\chi\@{ee}^{xx}}+{\chi\@{ee}^{yy}})}\Bigg] ,
    \end{split}
\end{equation}
where the top sign ($\pm$) corresponds to right-handed illumination, while the bottom sign describes left-handed transmission. The remaining six susceptibilities were characterized using the sampling points listed in Tab.~\ref{tab:split_ring_data} and displayed in Fig.~\ref{fig:babinet_split_ring_results}(c). The selection of the sampling points follows the ideas exposed in Sec.~\ref{sec:III_L_shape}, using normal incidence for characterizing in-plane susceptibilities (points 1 and 2), while out-of-plane susceptibilities were found using oblique incidence (points 3-6) with notable angles of incidence (which provides well-known trigonometric values). These sampling points also aim to provide diversity in terms of propagation planes, polarizations and scattering (transmission and reflection). The effective susceptibilities for this metasurface are also collected in Tab.~\ref{tab:split_ring_data}, with predicted co-polarized transmission and reflection displayed in  Fig.~\ref{fig:babinet_split_ring_results}(d). The complete set of scattering coefficients for simulation and predicted results are collected in the Supporting Information.

\begin{table}[h]
\centering
\caption{Sampling scenarios and normalized effective susceptibilities for the asymmetric split-ring metasurface of Fig.~\ref{fig:babinet_split_ring_results}.}
\begin{tabular}{ c  c  c  c }
\hline
& Parameter & $\theta$~[deg] & $\phi$~[deg] \\ [0.5ex] 
\hline
1& $T\@{TM,TM}$ & 0 & 0 \\
2& $T\@{TE,TE}$ & 0 & 0 \\
3& $T\@{TM,TM}$ & 15 & 0 \\
4& $R\@{TM,TM}$ & 30 & 90 \\
5& $T\@{TE,TE}$ & 15 & 90 \\
6& $R\@{TE,TE}$ & 30 & 0 \\
\hline
\end{tabular}
\quad
\begin{tabular}{ c  c }
\hline
& $\tilde{\chi}$ \\ [0.5ex] 
\hline
$\chi\@{ee}^{xx}$ & $-2.313 - 0.05715j$ \\
$\chi\@{ee}^{yy}$ & $-15.75 - 0.03367j$ \\
$\chi\@{ee}^{zz}$ & $0.01773 - 0.3191j$ \\
$\chi\@{mm}^{zz}$ & $-0.2511 - 1.844j$ \\
$\chi\@{em}^{yz}$ & $-7.806 + 13.83j$ \\
$\chi\@{em}^{zy}$ & $-3.178 + 4.102j$ \\
\hline
\end{tabular}
\label{tab:split_ring_data}
\end{table}

\begin{figure}[h]
\centering
\includegraphics[width=0.45\linewidth]{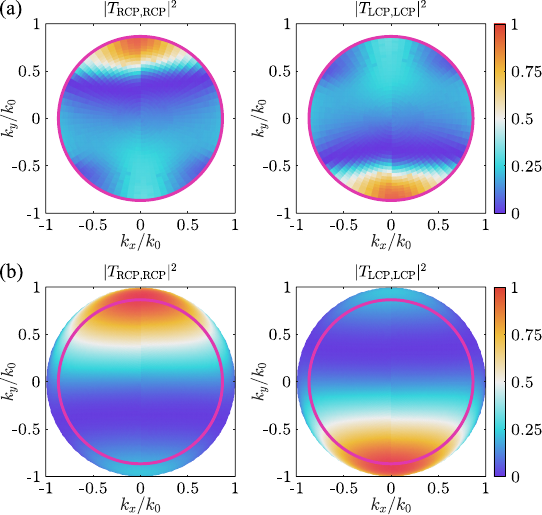}
\caption{Comparison between (a) the simulated data and (b) the model prediction considering co-polarized transmission for circular-polarized sources. The magenta line denotes the border of available simulated data for $\theta\leq60^{\circ}$.}\label{fig:babinet_split_ring_results_pt2}
\end{figure}

The comparison between the simulated results of Fig.~\ref{fig:babinet_split_ring_results}(c) with the prediction of Fig.~\ref{fig:babinet_split_ring_results}(d) reveals that most of the scattering coefficients match inside the simulated area inside the magenta circle with $\theta\leq60^{\circ}$. However, both reflection coefficients (and specially $R\@{TM,TM}$) diverge strongly between simulation and estimation for larger $\theta$ values, due to the excitation of high-order multipoles at such oblique angles. This divergence is also observed for circular-polarized waves, as shown in Fig.~\ref{fig:babinet_split_ring_results_pt2}, where transmission for right-hand polarized waves diverge between simulation and prediction for $k_y<0$ (and similarly for left-hand polarization considering scenarios with $k_y>0$). Nevertheless, the analytical model is able to demonstrate that the metasurface has extrinsic chirality, as it matches the metasurface behavior of allowing transmission of circular-polarized waves depending on the wave handiness and the illumination direction. Please note that the metasurface was characterized using sampling points from linear-polarized scattering, as shown in Fig.~\ref{fig:babinet_split_ring_results}(c), which are symmetric with respect $k_x$ and $k_y$.  

\subsection{Self-Complementary Zigzag Metasurfaces for Linear-to-Circular Polarization Conversion}\label{sec:III_zz_pattern}

\begin{figure}[h]
\centering
\includegraphics[width=1\linewidth]{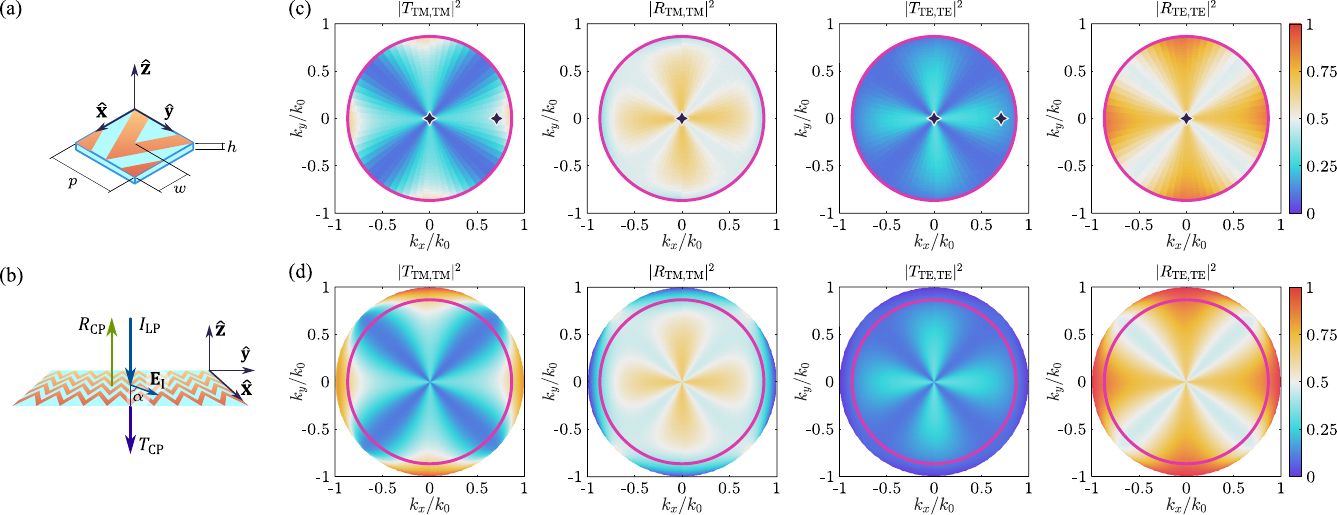}
\caption{(a) Schematic for the self-complementary zigzag metasurface unit-cell. (b) The metasurface is a linear-to-circular polarization converter when it is illuminated normally with a wave which electric fields points towards $\phi=\alpha$. (c) Simulation results and sampling points (denoted as dark stars) for linearly-polarized illumination. The magenta line show the upper bound for the simulation results at $\theta=60^{\circ}$. (d) Extrapolated results based on the effective susceptibilities calculated from the sampling points.   }\label{fig:zz_pattern_results}
\end{figure}

The last metasurface application is a self-complementary metasurface that converts linear-to-circular polarized converter, as portrayed in Fig.~\ref{fig:zz_pattern_results}(b). In detail, the metasurface converts a normal-incident source into circular-polarized scattering waves for specific polarization angles $\phi=\alpha=\pm 45^{\circ}$ \cite{Baena2017}. This example was selected precisely for its main feature, as most simulation software can provide information for either linear or circular polarizations, requiring post-processing to determine the polarization transformation.
The unit-cell of Fig.~\ref{fig:zz_pattern_results}(a) consider a copper pattern over an Arlon 25N substrate with $h=0.508$~[mm] thickness, while the pattern unit-cell has dimensions $p=6$~[mm] and $w=3.11$~[mm], using an operational frequency $f_0=8.665$~[GHz]~\cite{Baena2017}. By inspecting the pattern symmetries, the metasurface has two complementary unit-cells, one with $\sigma_y$ symmetry and the other with $C_{2,z}$. The reduced susceptibilities for such symmetries read
%
%
    \begin{equation}
        \dyad{\chi}\@{ee,mm}=
        \begin{bmatrix}
        \chi^{xx} & 0 & 0\\
        0 & \chi^{yy} & 0\\
        0 & 0 & \chi^{zz}
        \end{bmatrix} , \quad        \dyad{\chi}\@{em,me}=
        \begin{bmatrix}
        0 & \chi^{xy} & 0\\
        \chi^{yx} & 0 & 0\\
        0 & 0 & 0
        \end{bmatrix}. \label{eq:zz_pattern_susceptibilities}
    \end{equation}
With the susceptibilities of Eqs.~\eqref{eq:zz_pattern_susceptibilities}, and considering the reciprocity of the metasurface, it is possible to obtain linear-to-circular scattering parameters for oblique and normal incidence. As an example, the conversion from TM-polarized waves into right-hand circular transmission and reflection for a normal incidence scenario, but in-plane fields oriented in terms of an arbitrary $\phi$ angle read
\begin{subequations}
    \begin{equation}
    \begin{split}
    T\@{RCP,TM} = \frac{-j e^{j\phi}}{\sqrt{2} D\@{RCP,TM}}\bigg[&\sin (\phi) \left(k_0^2 {\chi\@{em}^{yx}}^2+4\right) \left(j k_0^2 \left({\chi\@{ee}^{xx}} {\chi\@{mm}^{yy}}+{\chi\@{em}^{xy}}^2\right)+2 k_0 ({\chi\@{ee}^{xx}}+{\chi\@{mm}^{yy}})-4 j\right)\\
    &-\cos (\phi) \left(k_0^2 {\chi\@{em}^{yx}}^2-2 j k_0 {\chi\@{ee}^{yy}}-4\right) \left(k_0^2 \left({\chi\@{ee}^{xx}} {\chi\@{mm}^{yy}}+{\chi\@{em}^{xy}}^2\right)+4\right)\bigg] ,
    \end{split}
    \end{equation}
    \begin{equation}
    \begin{split}
    R\@{RCP,TM} = -\frac{\sqrt{2} k_0 e^{-j\phi}}{D\@{RCP,TM}} \bigg[&j \sin (\phi) ({\chi\@{ee}^{yy}}-2 {\chi\@{em}^{yx}}) \left(k_0^2 \left({\chi\@{ee}^{xx}} {\chi\@{mm}^{yy}}+{\chi\@{em}^{xy}}^2\right)-2 j k_0 ({\chi\@{ee}^{xx}}+{\chi\@{mm}^{yy}})-4\right)\\
    &+\left(\cos (\phi) \left(k_0^2 {\chi\@{em}^{yx}}^2-2 j k_0 {\chi\@{ee}^{yy}}-4\right) ({\chi\@{ee}^{xx}}+2 {\chi\@{em}^{xy}}-{\chi\@{mm}^{yy}})\right)\bigg] ,
    \end{split}
    \end{equation}
    \begin{equation}
    D\@{RCP,TM} = \left(k_0^2 {\chi\@{em}^{yx}}^2-2 j k_0 {\chi\@{ee}^{yy}}-4\right) \left(k_0^2 \left({\chi\@{ee}^{xx}} {\chi\@{mm}^{yy}}+{\chi\@{em}^{xy}}^2\right)-2 j k_0 ({\chi\@{ee}^{xx}}+{\chi\@{mm}^{yy}})-4\right)     .
    \end{equation}
\end{subequations}

Following the same procedure for the previous examples [using a Mathematica model implementation to determine analytical expressions for the susceptibilities of Eqs.~\eqref{eq:zz_pattern_susceptibilities} and the MATLAB \verb+vpasolve+ for a set of illumination samples], it was found that the susceptibilities $\chi\@{ee}^{zz}$ and $\chi\@{mm}^{xx}$ are negligible compared to the rest of susceptibilities and, therefore, can be discarded in order to reduce the characterization complexity. The metasurface was characterized using co-polarized transmission and reflection at normal incidence, and only co-polarized transmission for $\theta=45^{\circ}$, $\phi=0^{\circ}$ with both linear polarizations, as summarized in Tab.~\ref{tab:zz_pattern_data} and shown in Fig.~\ref{fig:zz_pattern_results}(c). As done with previous examples, the normalized susceptibilities are also summarized in Tab.~\ref{tab:zz_pattern_data}.

\begin{table}[h]
\centering
\caption{Sampling scenarios and normalized effective susceptibilities for the self-complementary zigzag metasurface of Fig.~\ref{fig:zz_pattern_results}.}
\begin{tabular}{ c  c  c  c }
\hline
& Parameter & $\theta$~[deg] & $\phi$~[deg] \\ [0.5ex] 
\hline
1& $T\@{TM,TM}$ & 0 & 0 \\
2& $T\@{TE,TE}$ & 0 & 0 \\
3& $R\@{TM,TM}$ & 0 & 0 \\
4& $R\@{TE,TE}$ & 0 & 0 \\
5& $T\@{TM,TM}$ & 45 & 0 \\
6& $T\@{TE,TE}$ & 45 & 0 \\
\hline
\end{tabular}
\quad
\begin{tabular}{ c  c }
\hline
& $\tilde{\chi}$ \\ [0.5ex] 
\hline
$\chi\@{ee}^{xx}$ & $2.905 - 0.479j$ \\
$\chi\@{ee}^{yy}$ & $-2.908 - 0.0102j$ \\
$\chi\@{mm}^{yy}$ & $0.1515 - 0.0349j$ \\
$\chi\@{mm}^{zz}$ & $0.1779 + 0.0013j$ \\
$\chi\@{em}^{xy}$ & $0.2026 + 0.1715j$ \\
$\chi\@{em}^{yx}$ & $0.0006 + 0.01j$ \\
\hline
\end{tabular}
\label{tab:zz_pattern_data}
\end{table}

\begin{figure}[h]
\centering
\includegraphics[width=1\linewidth]{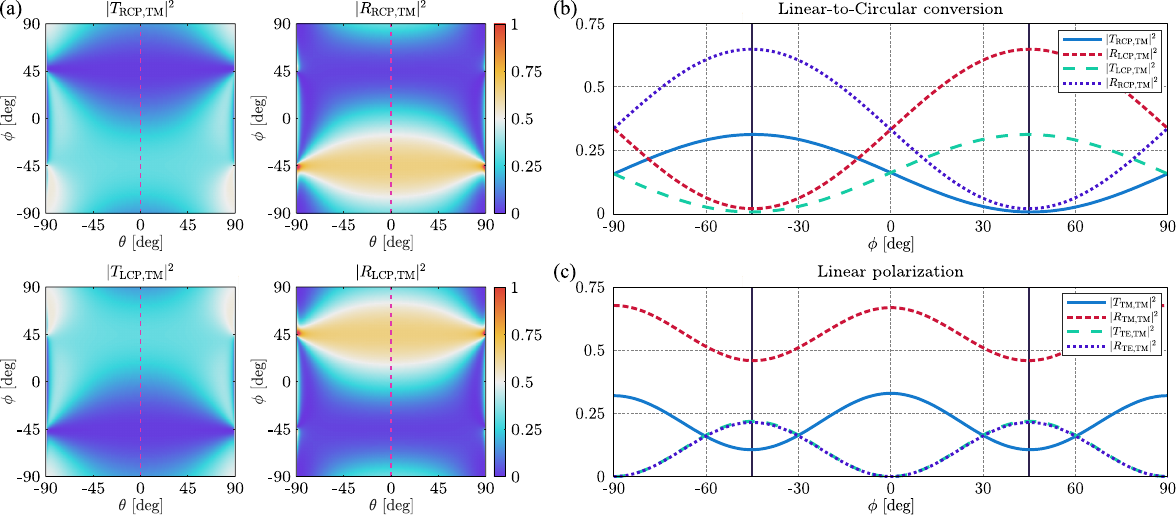}
\caption{(a) Conversion from TM-polarized waves into left- and right-hand circular polarized waves for oblique incidence. The dotted lines denotes scattering under normal incidence. (b) Linear-to-circular polarization conversion for TM incident waves as a function of $\phi$, for the case when $\theta=0$. (c) Co- and cross-polarized scattering for normal-incidence scenario as a function of $\phi$ with TM waves. The dark lines denotes the case with incident polarization $\phi=\alpha=45^{\circ}.$}\label{fig:zz_pattern_results_pt2}
\end{figure}

Unlike the previous examples, the self complementary metasurface shows a mostly-dipole response as the predicted analytical values of Fig.~\ref{fig:zz_pattern_results}(d) matches with the simulated data from Fig.~\ref{fig:zz_pattern_results}(c) for almost every angle of incidence. 
In terms of linear-to-circular polarization, Fig.~\ref{fig:zz_pattern_results_pt2}(a) shows that the metasurface is able to convert TM polarized waves into left-hand polarized waves for incident angles $\phi=45^{\circ}$, almost independent of the value of $\theta$; while the conversion into right-hand polarized waves occurs for $\phi=-45^{\circ}$. By observing Fig.~\ref{fig:zz_pattern_results_pt2}(b), it can be found out that, as mentioned in Ref.~\cite{Baena2017}, for in-plane angle of incidence $\phi=45^{\circ}$ the metasurface transform linearly-polarized incident light into circular-polarized ones, with the handiness depending on the value of $\phi$. In terms of linear polarization, as shown in Fig.~\ref{fig:zz_pattern_results_pt2}(c),  it can be see that in cases when the electric field is aligned with the metasurface mesh coordinates (either $x$ or $y$), there is weak cross-polarized response, while it is maximized for $\phi=\pm \alpha$. This results in a weak circular-to-linear conversion for cases near $\phi=0^{\circ},90^{\circ}$, while the intended functionality is achieved at $\phi=45^{\circ}$.

\section{Conclusion}\label{sec:conclusion}

In this work, we have presented a generalized solution for the GSTCs considering a uniform metasurface homogenized in terms of susceptibilities. Thanks to the model developed in this work, it is possible to determine the specular scattering (transmission and reflection) for a metasurface illuminated from an arbitrary direction, regardless of the source linear polarization (by providing co- and cross-polarized scattering terms). By solving the GSTCs in a generic scenario, given the susceptibility tensors, it is possible to obtain analytical scattering expressions without having to solve a system of equations with variable complexity. Instead, the provided model can be implemented numerically to be solved using $2\times 2$ matrices. By exploiting the symmetries of the unit-cell, it is possible to reduce the analytical and computational complexity of the scattering coefficients by decreasing the number of susceptibilities required to describe the metasurface. 

In addition, this work presented three different scenarios where the model is not only used to obtain analytical expressions for transmission and reflection coefficients, but also to characterize each metasurface susceptibilities by selecting appropriate illumination cases. 
In the first example, it was shown that the model can replicate the behavior of a metasurface located between two different media using the susceptibilities provided in Ref.~\cite{Tiukuvaara2022}. The second example proved that, exploiting the unit-cell symmetries and using samples for linearly-polarized cases, it is possible to characterize a metasurface with extrinsic chirality under circular-polarized illumination. The final example demonstrated capabilities of the model to characterize hybrid-polarization scattering (linear-to-circular polarization) for normal and oblique-incidence illumination. In summary, the model proposed in this work is a powerful tool for metasurface design and characterization as it can determine the specular scattering of a metasurface given its susceptibilities and vice-versa. 


\medskip
\textbf{Acknowledgements} \par 
We acknowledge funding from the Swiss National Science Foundation (project TMSGI2\_218392).

\medskip

%

\bibliographystyle{IEEEtran}
\bibliography{references}

\newpage

\renewcommand{\theequation}{S\arabic{equation}}
\renewcommand{\thefigure}{S\arabic{figure}}
\renewcommand{\thesection}{S\arabic{section}}
\renewcommand{\thesubsection}{\thesection.\arabic{subsection}}
\setcounter{section}{0}
\setcounter{equation}{0}
 

\newpage
\part*{Supporting Information}
\section{Example of model implementation in MATLAB}
Code also available at \url{https://doi.org/10.5281/zenodo.14929637}.
\begin{lstlisting}[style=Matlab-editor]
%%%constants setup
eps0 = 8.85418e-12;
mu0 = 1.256637e-6;
c0 = 1/sqrt(eps0*mu0);
eta0 = sqrt(mu0/eps0);

%%%frequency (arbitrary) setup
f0 = 8.665e9;
%l0 = 655e-9;
l0=c0/f0;
k0 = 2*pi/l0;
w0 = k0*c0;
%f0 = w0/(2*pi);

%%media setup
er1 = 1;
mr1 = 1;

er2 = 1;
mr2 = 1;

eps1 = er1*eps0;
mu1 = mr1*mu0;

eps2 = er2*eps0;
mu2 = mr2*mu0;

k1 = k0*sqrt(er1*mr1);
k2 = k0*sqrt(er2*mr2);

eta1 = sqrt(mu1/eps1);
eta2 = sqrt(mu2/eps2);

%%%Direction of illumination: 0->fwd(u->d), 1->bckwd(d->u)
dirsw=0;%<-choose here
dirsgn=(-1)^dirsw;%<-sign switch
dirsw=(1-dirsgn)/2;%<-binary buffer

%%%Direction of normal vector with respect to z axis
dirn=-1;

%%%setup angle of incidence
% thetaspn = [0, 1e-3, 1:89];%sweep in terms of theta
% phispn = 0:360;%sweep in terms of phi

thetaspn = [-89:-1, -1e-3,0, 1e-3, 1:89];%sweep in terms of theta
phispn = -90:90;%sweep in terms of phi

[TT,PP] = meshgrid(thetaspn,phispn);

%%%Susceptibilities initialization
CChiee = zeros(3);
CChimm = zeros(3);
CChiem = zeros(3);
CChime = zeros(3);

CChiee(1,1)=(2.9047 - 0.4790i)/k0;
CChiee(2,2)=(-2.9079 - 0.0102i)/k0;
CChiee(3,3)=(0)/k0;

CChimm(1,1)=(0)/k0;
CChimm(2,2)=(0.1515 - 0.0349i)/k0;
CChimm(3,3)=(0.1779 + 0.0013i)/k0;

CChiem(1,2)=(0.2026 + 0.1715i)/k0;
CChiem(2,1)=( 0.0006 + 0.0100i)/k0;
           

%%%Code to impose reciprocity over the susceptibilities
CChiee = CChiee + transpose(CChiee) - eye(3).*CChiee;
CChimm = CChimm + transpose(CChimm) - eye(3).*CChimm;
CChiemtemp = CChiem - transpose(CChime);
CChiem = CChiemtemp;
CChime = - transpose(CChiemtemp);

%%Susceptibilities decomposition into tangential and normal components
CChiee_tt = eps0*CChiee(1:2,1:2);
Chiee_tn = dirn*eps0*CChiee(1:2,3);
Chiee_nt = dirn*eps0*transpose(CChiee(3,1:2));
chiee_nn = eps0*CChiee(3,3);

CChiem_tt = (1/c0)*CChiem(1:2,1:2);
Chiem_tn = dirn*(1/c0)*CChiem(1:2,3);
Chiem_nt = dirn*(1/c0)*transpose(CChiem(3,1:2));
chiem_nn = (1/c0)*CChiem(3,3);

CChime_tt = (1/c0)*CChime(1:2,1:2);
Chime_tn = dirn*(1/c0)*CChime(1:2,3);
Chime_nt = dirn*(1/c0)*transpose(CChime(3,1:2));
chime_nn = (1/c0)*CChime(3,3);

CChimm_tt = mu0*CChimm(1:2,1:2);
Chimm_tn = dirn*mu0*CChimm(1:2,3);
Chimm_nt = dirn*mu0*transpose(CChimm(3,1:2));
chimm_nn = mu0*CChimm(3,3);

%%%Other required matrices
IIt = eye(2);
NCross =  dirn*[0, - 1;1,0];% for operator Nx(something)

%%%Transformation matrices for circular-polarization
UU=[1j,-1j;1,1]/sqrt(2);
UUt=UU.';
UUi=UU\IIt;

%%%Sweep in terms of theta
%%co-polarized scattering coefficients (linear polarization)
tautete = zeros(size(TT));
tautmtm = zeros(size(TT));
gammatete = zeros(size(TT));
gammatmtm = zeros(size(TT));

%%cross-polarized scattering coefficients (linear polarization)
tautetm = zeros(size(TT));
tautmte = zeros(size(TT));
gammatetm = zeros(size(TT));
gammatmte = zeros(size(TT));

%%co-polarized scattering coefficients (circular polarization)
taurcrc = zeros(size(TT));
taulclc = zeros(size(TT));
gammarcrc = zeros(size(TT));
gammalclc = zeros(size(TT));

%%cross-polarized scattering coefficients (circular polarization)
taurclc = zeros(size(TT));
taulcrc = zeros(size(TT));
gammarclc = zeros(size(TT));
gammalcrc = zeros(size(TT));

%%linear-to-circular polarization scattering coefficients
taurhtm = zeros(size(TT));
taulhte = zeros(size(TT));
gammarhtm = zeros(size(TT));
gammalhte = zeros(size(TT));

taurhte = zeros(size(TT));
taulhtm = zeros(size(TT));
gammarhte = zeros(size(TT));
gammalhtm = zeros(size(TT));

for mth = 1:length(phispn)
    phi=phispn(mth);

%%Defining unitary tangential vectors
    Ukt = [cosd(phi);sind(phi)];
    Ult = NCross*Ukt;

%%Defining unitary linear polarization vectors    
    Qte = -Ult;
    Qtm = Ukt;

    for nth = 1:length(thetaspn)
    %%angle and wavevector setup
        theta = thetaspn(nth);

        kt = k1*sind(theta);
        Kt = kt*Ukt;
        Lt = NCross*Kt;
        
        kz1 = sqrt(abs(k1).^2 - kt.^2);
        kz2 = sqrt(abs(k2).^2 - kt.^2);
      
    %%Tensor combinations
        kt_kt = tensorbuild(Ukt,Ukt);
        nkt_nkt = tensorbuild(Ult,Ult);
    
    %%Wave impedances
        ztm1 = eta1*sqrt(1 - (kt/k1).^2);
        zte1 = eta1/sqrt(1 - (kt/k1).^2);
        ztm2 = eta2*sqrt(1 - (kt/k2).^2);
        zte2 = eta2/sqrt(1 - (kt/k2).^2);
        
        ztmu = (1-dirsw)*ztm1 + dirsw*ztm2;
        zteu = (1-dirsw)*zte1 + dirsw*zte2;
        ztmd = (1-dirsw)*ztm2 + dirsw*ztm1;
        zted = (1-dirsw)*zte2 + dirsw*zte1;
        
        ZZu = ztmu*kt_kt + zteu*nkt_nkt;
        ZZd = ztmd*kt_kt + zted*nkt_nkt;
        YYu = ZZu\IIt;
        YYd = ZZd\IIt;

        ZZusqrt = sqrt(ztmu)*kt_kt + sqrt(zteu)*nkt_nkt;
        ZZdsqrt = sqrt(ztmd)*kt_kt + sqrt(zted)*nkt_nkt;
        YYusqrt = ZZusqrt\IIt;
        YYdsqrt = ZZdsqrt\IIt;
    
    %%Scattering fields calculation, for forward illumination case
    
        KtYYu = Kt/ztmu;
        KtYYd = Kt/ztmd;
    
        VV_ti = (CChiee_tt - dirsgn*CChiem_tt*(NCross*YYu) + ...
                    dirsgn*tensorbuild(Chiee_tn,KtYYu)/(w0*eps0) + ...
                    tensorbuild(Chiem_tn,Lt)/(w0*mu0))/2;
        V_ni = (Chiee_nt - dirsgn*transpose(transpose(Chiem_nt)*(NCross*YYu)) + ...
                    dirsgn*(chiee_nn)/(w0*eps0)*KtYYu + ...
                    (chiem_nn)/(w0*mu0)*(Lt))/2;
        WW_ti = (CChime_tt - dirsgn*CChimm_tt*(NCross*YYu) + ...
                    dirsgn*tensorbuild(Chime_tn,KtYYu)/(w0*eps0) + ...
                    tensorbuild(Chimm_tn,Lt)/(w0*mu0))/2;
        W_ni = (Chime_nt - dirsgn*transpose(transpose(Chimm_nt)*(NCross*YYu)) + ...
                    dirsgn*(chime_nn)/(w0*eps0)*KtYYu + ...
                    (chimm_nn)/(w0*mu0)*(Lt))/2;
        
        VV_tr = (CChiee_tt + dirsgn*CChiem_tt*(NCross*YYu) + ...
                    (-1)*dirsgn*tensorbuild(Chiee_tn,KtYYu)/(w0*eps0) + ...
                    tensorbuild(Chiem_tn,Lt)/(w0*mu0))/2;
        V_nr = (Chiee_nt + dirsgn*transpose(transpose(Chiem_nt)*(NCross*YYu)) + ...
                    (-1)*dirsgn*(chiee_nn)/(w0*eps0)*KtYYu + ...
                    (chiem_nn)/(w0*mu0)*(Lt))/2;
        WW_tr = (CChime_tt + dirsgn*CChimm_tt*(NCross*YYu) + ...
                    (-1)*dirsgn*tensorbuild(Chime_tn,KtYYu)/(w0*eps0) + ...
                    tensorbuild(Chimm_tn,Lt)/(w0*mu0))/2;
        W_nr = (Chime_nt + dirsgn*transpose(transpose(Chimm_nt)*(NCross*YYu)) + ...
                    (-1)*dirsgn*(chime_nn)/(w0*eps0)*KtYYu + ...
                    (chimm_nn)/(w0*mu0)*(Lt))/2;
        
        VV_tt = (CChiee_tt - dirsgn*CChiem_tt*(NCross*YYd) + ...
                    dirsgn*tensorbuild(Chiee_tn,KtYYd)/(w0*eps0) + ...
                    tensorbuild(Chiem_tn,Lt)/(w0*mu0))/2;
        V_nt = (Chiee_nt - dirsgn*transpose(transpose(Chiem_nt)*(NCross*YYd)) + ...
                    dirsgn*(chiee_nn)/(w0*eps0)*KtYYd + ...
                    (chiem_nn)/(w0*mu0)*(Lt))/2;
        WW_tt = (CChime_tt - dirsgn*CChimm_tt*(NCross*YYd) + ...
                    dirsgn*tensorbuild(Chime_tn,KtYYd)/(w0*eps0) + ...
                    tensorbuild(Chimm_tn,Lt)/(w0*mu0))/2;
        W_nt = (Chime_nt - dirsgn*transpose(transpose(Chimm_nt)*(NCross*YYd)) + ...
                    dirsgn*(chime_nn)/(w0*eps0)*KtYYd + ...
                    (chimm_nn)/(w0*mu0)*(Lt))/2;
        
  
    
    %%Replacing surface polarizations into the GSTC
        BBei = +dirsgn*IIt - 1j*(w0*NCross*WW_ti + tensorbuild(Kt,V_ni)/eps0);
        BBer = +dirsgn*IIt - 1j*(w0*NCross*WW_tr + tensorbuild(Kt,V_nr)/eps0);
        BBet = -dirsgn*IIt - 1j*(w0*NCross*WW_tt + tensorbuild(Kt,V_nt)/eps0);
    
        BBhi = +YYu - 1j*(w0*VV_ti + tensorbuild(Lt,W_ni)/mu0);
        BBhr = -YYu - 1j*(w0*VV_tr + tensorbuild(Lt,W_nr)/mu0);
        BBht = -YYd - 1j*(w0*VV_tt + tensorbuild(Lt,W_nt)/mu0);    
    
	%%Solution to the GSTC
        AAr1 = BBer - BBet*(BBht\BBhr);
        AAr2 = BBei - BBet*(BBht\BBhi);
        GGammatg = - AAr1\AAr2;
  
        AAt1 = BBet - BBer*(BBhr\BBht);
        AAt2 = BBei - BBer*(BBhr\BBhi);
        TTautg = - AAt1\AAt2;
    
    %%Conversion from tangential to normalized fields
        GGamma = YYusqrt*GGammatg*ZZusqrt;
        TTau = YYdsqrt*TTautg*ZZusqrt;

        Tau_lin = zeros(2);
        Gamma_lin = zeros(2);

    %%Scattering coefficient extraction, using tensor dot product
        Tau_lin(2,2) = dot(Qte,TTau*Qte);%TETE
        Tau_lin(1,1) = dot(Qtm,TTau*Qtm);%TMTM
        Gamma_lin(2,2) = dot(Qte,GGamma*Qte);%TETE
        Gamma_lin(1,1) = dot(Qtm,GGamma*Qtm);%TMTM

        Tau_lin(1,2) =dot(Qtm,TTau*Qte);%TMTE
        Tau_lin(2,1) = dot(Qte,TTau*Qtm);%TETM
        Gamma_lin(1,2) = dot(Qtm,GGamma*Qte);%TMTE
        Gamma_lin(2,1) = dot(Qte,GGamma*Qtm);%TETM

        %%Storing linear-polarization results
        tautete(mth,nth) = Tau_lin(2,2);
        tautmtm(mth,nth) = Tau_lin(1,1);
        gammatete(mth,nth) = Gamma_lin(2,2);
        gammatmtm(mth,nth) = Gamma_lin(1,1);

        tautmte(mth,nth) = Tau_lin(1,2);
        tautetm(mth,nth) = Tau_lin(2,1);
        gammatmte(mth,nth) = Gamma_lin(1,2);
        gammatetm(mth,nth) = Gamma_lin(2,1);

%%%%%%%%%%%%%%%%%%%%%%%%%%%%%%%%%%%%%%%%%%%%%%%%%%%%%%%%%%%%%%%%%%%%%%%%%%%
%% Calculating circular-polarization coefficients
%%%%%%%%%%%%%%%%%%%%%%%%%%%%%%%%%%%%%%%%%%%%%%%%%%%%%%%%%%%%%%%%%%%%%%%%%%%               

        Tau_circ=UUi*(Tau_lin*UU);
        Gamma_circ=UUt*(Gamma_lin*UU);
        
        %%Storing circular-polarization results
        taurcrc(mth,nth) = Tau_circ(1,1);
        taulclc(mth,nth) = Tau_circ(2,2);
        gammarcrc(mth,nth) = Gamma_circ(1,1);
        gammalclc(mth,nth) = Gamma_circ(2,2);
        
        taurclc(mth,nth) = Tau_circ(1,2);
        taulcrc(mth,nth) = Tau_circ(2,1);
        gammarclc(mth,nth) = Gamma_circ(1,2);
        gammalcrc(mth,nth) = Gamma_circ(2,1);
    
%%%%%%%%%%%%%%%%%%%%%%%%%%%%%%%%%%%%%%%%%%%%%%%%%%%%%%%%%%%%%%%%%%%%%%%%%%%
%% Calculating linear-to-circular polarization coefficients
%%%%%%%%%%%%%%%%%%%%%%%%%%%%%%%%%%%%%%%%%%%%%%%%%%%%%%%%%%%%%%%%%%%%%%%%%%%               

        Tau_cplp=UUi*(Tau_lin);
        Gamma_cplp=UUt*(Gamma_lin);

        %%Storing circular-polarization results
        taurhtm(mth,nth) = Tau_cplp(1,1);
        taulhte(mth,nth) = Tau_cplp(2,2);
        gammarhtm(mth,nth) = Gamma_cplp(1,1);
        gammalhte(mth,nth) = Gamma_cplp(2,2);

        taurhte(mth,nth) = Tau_cplp(1,2);
        taulhtm(mth,nth) = Tau_cplp(2,1);
        gammarhte(mth,nth) = Gamma_cplp(1,2);
        gammalhtm(mth,nth) = Gamma_cplp(2,1);

    end
end

%%%%%%%%%%%%%%%%%%%%%%%%%%%%%%%%%%%%%%%%%%%%%%%%%%%%%%%%%%%%%%%%%%%%%%%%%%%
%%%Additional function to create tensors from single vectors
%%%%%%%%%%%%%%%%%%%%%%%%%%%%%%%%%%%%%%%%%%%%%%%%%%%%%%%%%%%%%%%%%%%%%%%%%%%

function TT = tensorbuild(A,B)
TT = A*transpose(B);
end
\end{lstlisting}

\newpage
\section{Simulation and prediction plots}
\subsection{Transmissive Angular-Asymmetric Metasurface}

\begin{figure}[!h]
\centering
\includegraphics[width=1\linewidth]{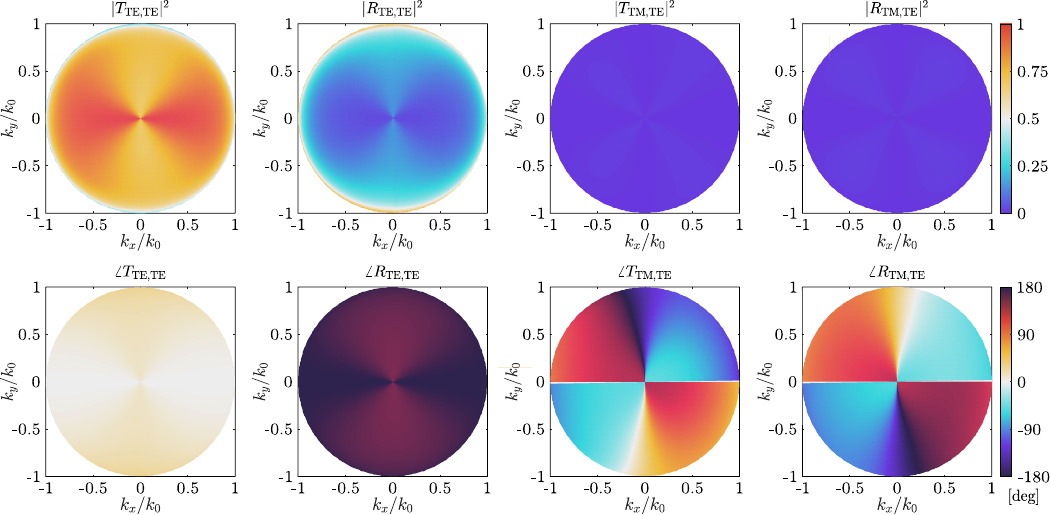}
\caption{Predicted scattering under TE illumination}\label{fig:mod_l_shape_lp_te}
\end{figure}

\begin{figure}[!h]
\centering
\includegraphics[width=1\linewidth]{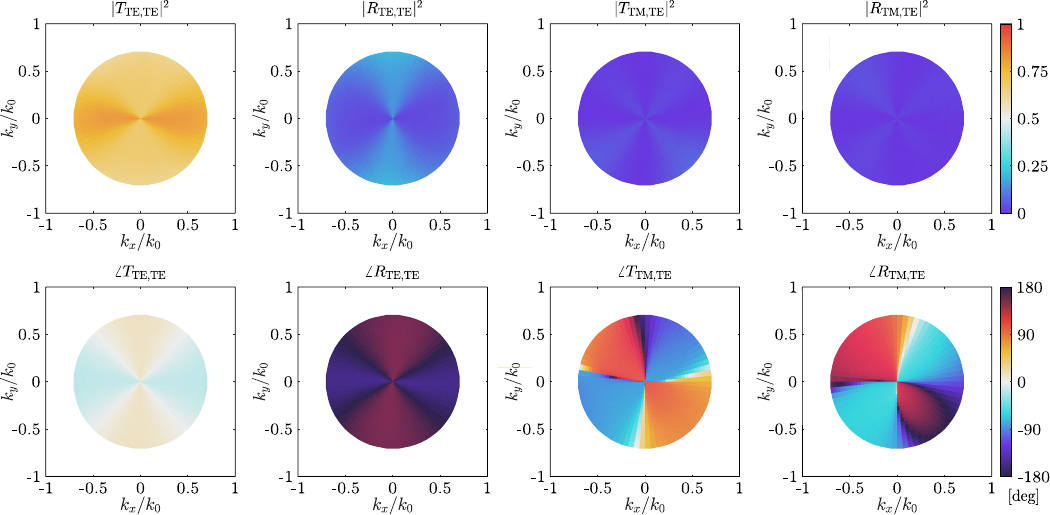}
\caption{Simulation results for TE illumination}\label{fig:sim_l_shape_lp_te}
\end{figure}

\newpage

\begin{figure}[!h]
\centering
\includegraphics[width=1\linewidth]{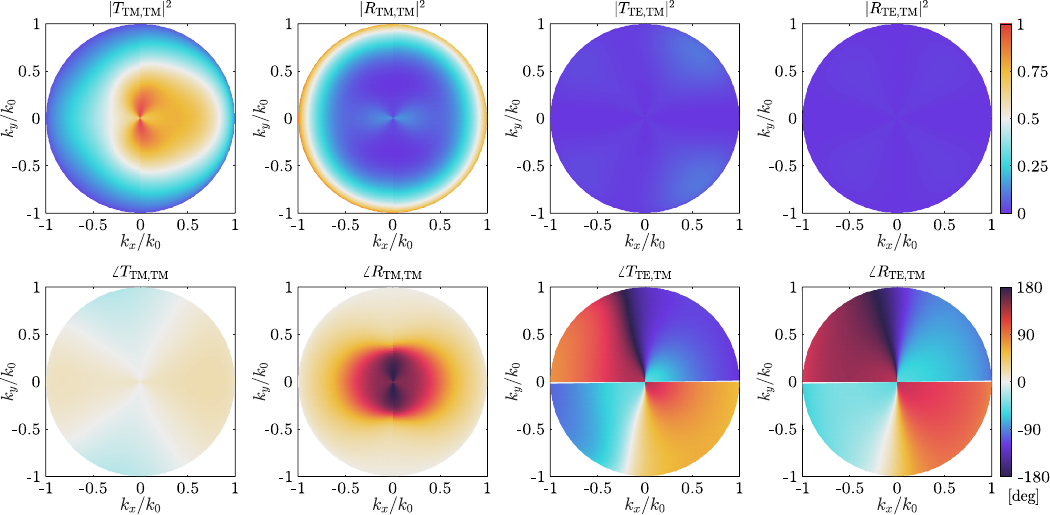}
\caption{Predicted scattering under TM illumination}\label{fig:mod_l_shape_lp_tm}
\end{figure}

\begin{figure}[!h]
\centering
\includegraphics[width=1\linewidth]{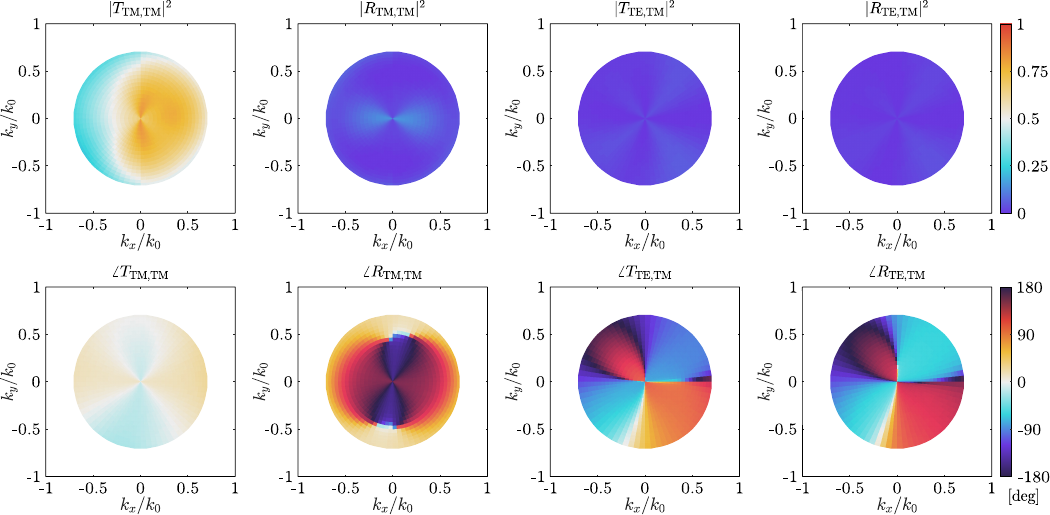}
\caption{Simulation results for TM illumination}\label{fig:sim_l_shape_lp_tm}
\end{figure}

\newpage
\subsection{Asymmetric Achiral Split-Ring with Optical Activity}

\begin{figure}[!h]
\centering
\includegraphics[width=1\linewidth]{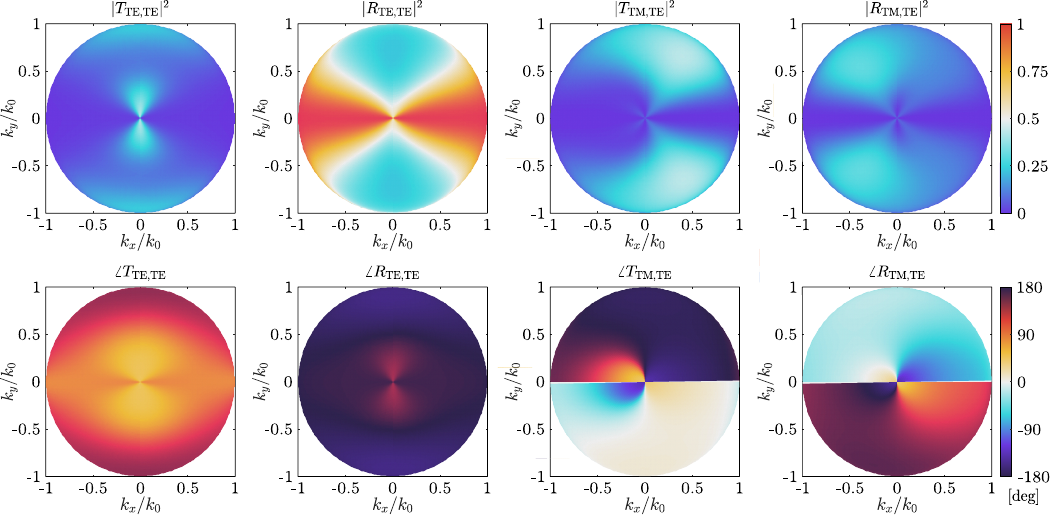}
\caption{Predicted scattering under TE illumination}\label{fig:mod_babinet_split_ring_lp_te}
\end{figure}

\begin{figure}[!h]
\centering
\includegraphics[width=0.55\linewidth]{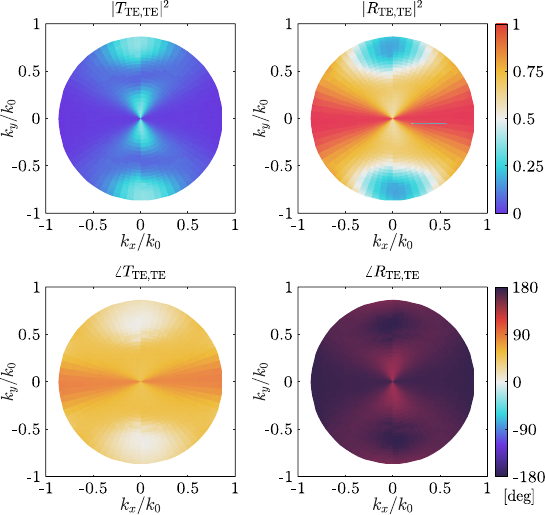}
\caption{Simulation results for TE illumination (co-polarization only)}\label{fig:sim_babinet_split_ring_lp_te}
\end{figure}

\newpage

\begin{figure}[!h]
\centering
\includegraphics[width=1\linewidth]{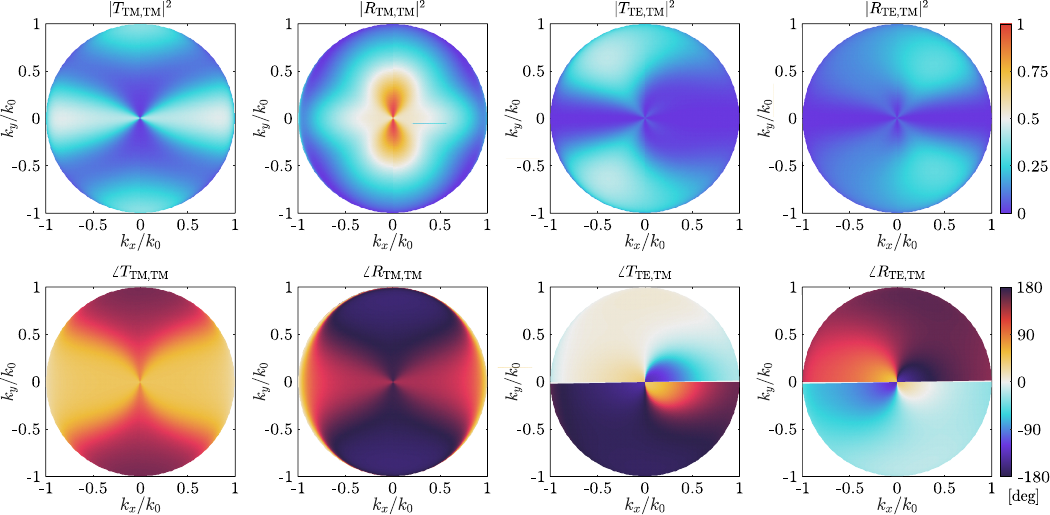}
\caption{Predicted scattering under TM illumination}\label{fig:mod_babinet_split_ring_lp_tm}
\end{figure}

\begin{figure}[!h]
\centering
\includegraphics[width=0.55\linewidth]{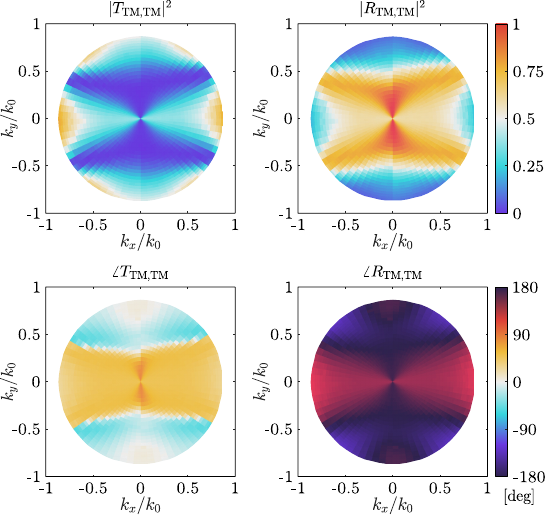}
\caption{Simulation results for TM illumination (co-polarization only)}\label{fig:sim_babinet_split_ring_lp_tm}
\end{figure}

\newpage

\begin{figure}[!h]
\centering
\includegraphics[width=1\linewidth]{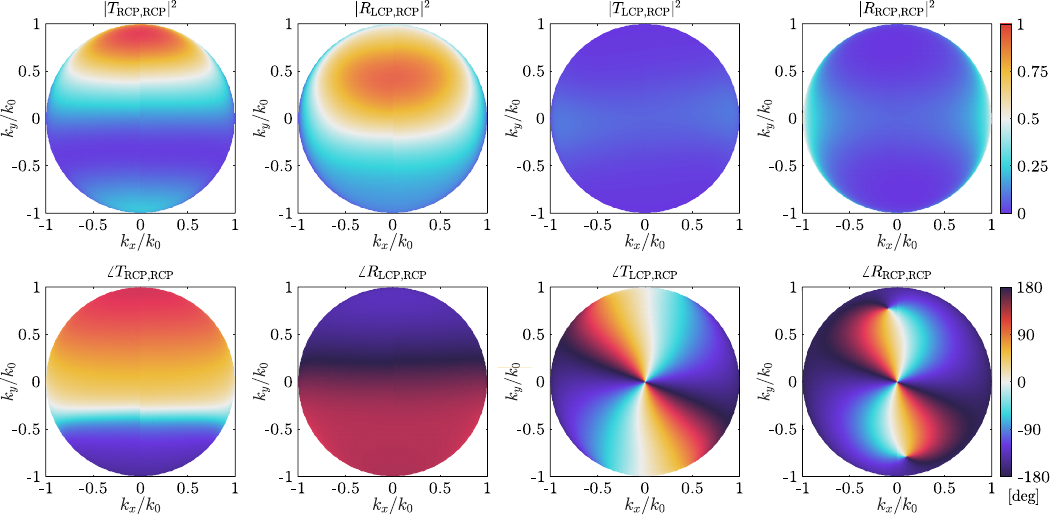}
\caption{Predicted scattering under RCP illumination}\label{fig:mod_babinet_split_ring_cp_rcp}
\end{figure}

\begin{figure}[!h]
\centering
\includegraphics[width=0.55\linewidth]{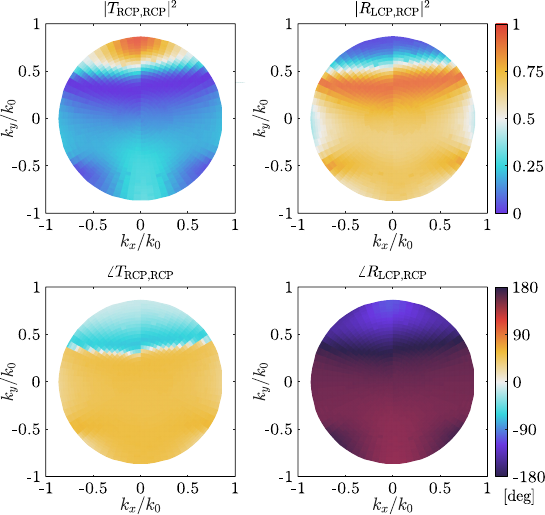}
\caption{Simulation results for RCP illumination (co-polarization only)}\label{fig:sim_babinet_split_ring_cp_rcp}
\end{figure}

\newpage

\begin{figure}[!h]
\centering
\includegraphics[width=1\linewidth]{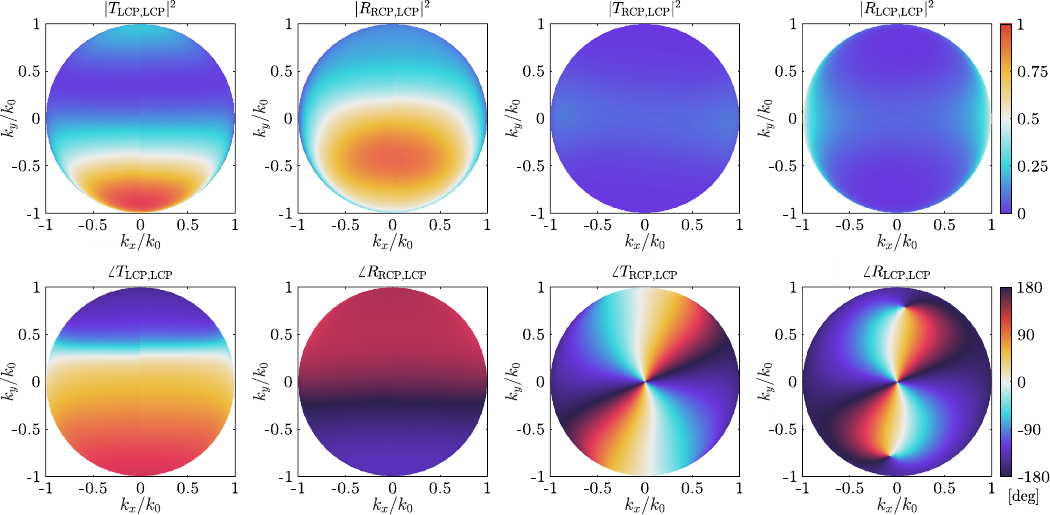}
\caption{Predicted scattering under LCP illumination}\label{fig:mod_babinet_split_ring_cp_lcp}
\end{figure}

\begin{figure}[!h]
\centering
\includegraphics[width=0.55\linewidth]{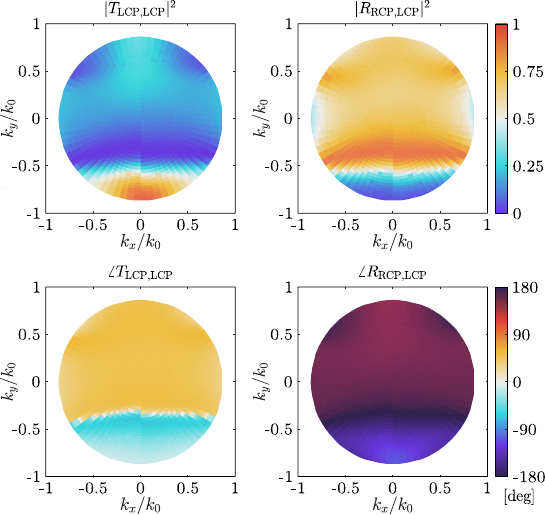}
\caption{Simulation results for LCP illumination (co-polarization only)}\label{fig:sim_babinet_split_ring_cp_lcp}
\end{figure}

\newpage
\subsection{Self-Complementary Zigzag Metasurfaces for Linear-to-Circular Polarization Conversion}

\begin{figure}[!h]
\centering
\includegraphics[width=1\linewidth]{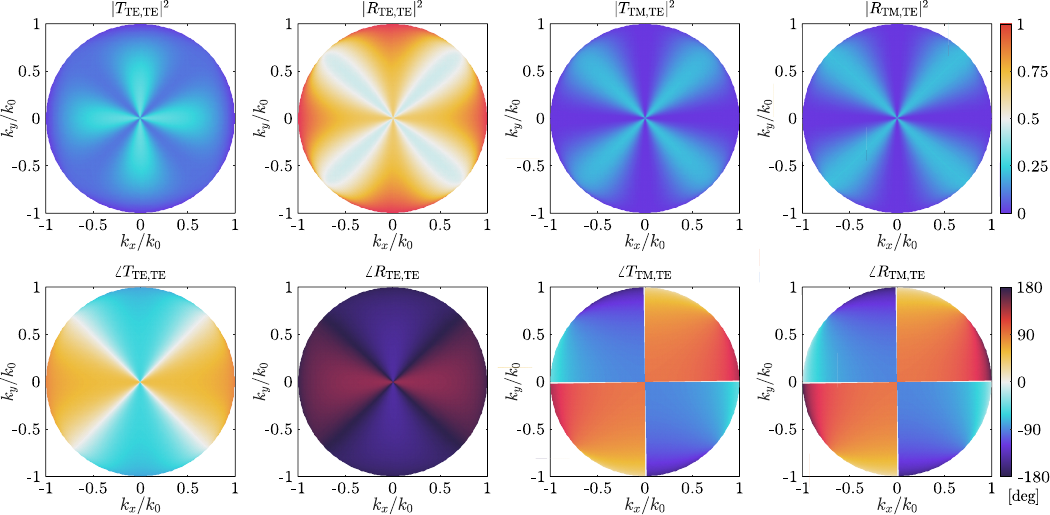}
\caption{Predicted scattering under TE illumination}\label{fig:mod_zz_pattern_lp_te}
\end{figure}

\begin{figure}[!h]
\centering
\includegraphics[width=1\linewidth]{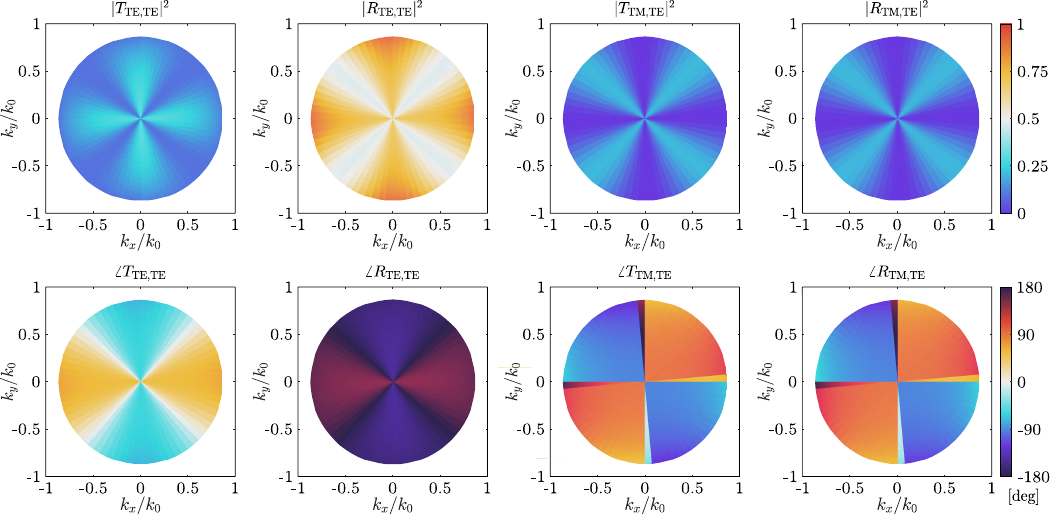}
\caption{Simulation results for TE illumination}\label{fig:sim_zz_pattern_lp_te}
\end{figure}

\newpage

\begin{figure}[!h]
\centering
\includegraphics[width=1\linewidth]{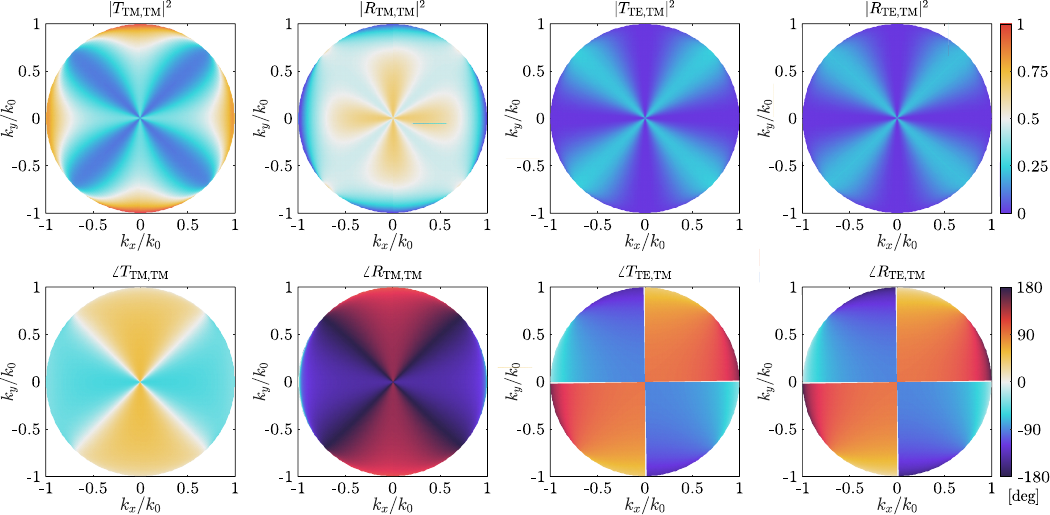}
\caption{Predicted scattering under TM illumination}\label{fig:mod_zz_pattern_lp_tm}
\end{figure}

\begin{figure}[!h]
\centering
\includegraphics[width=1\linewidth]{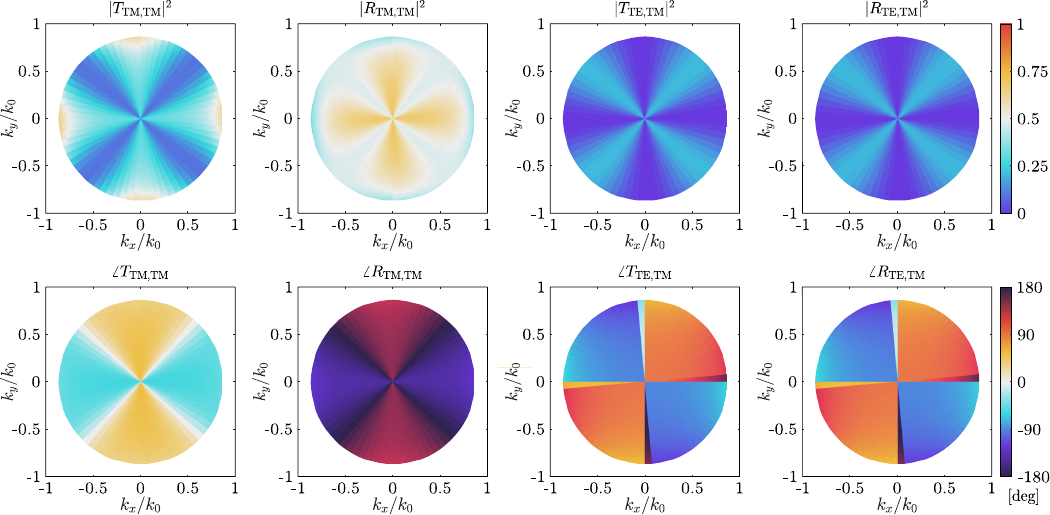}
\caption{Simulation results for TM illumination}\label{fig:sim_zz_pattern_lp_tm}
\end{figure}

\newpage

\begin{figure}[!h]
\centering
\includegraphics[width=1\linewidth]{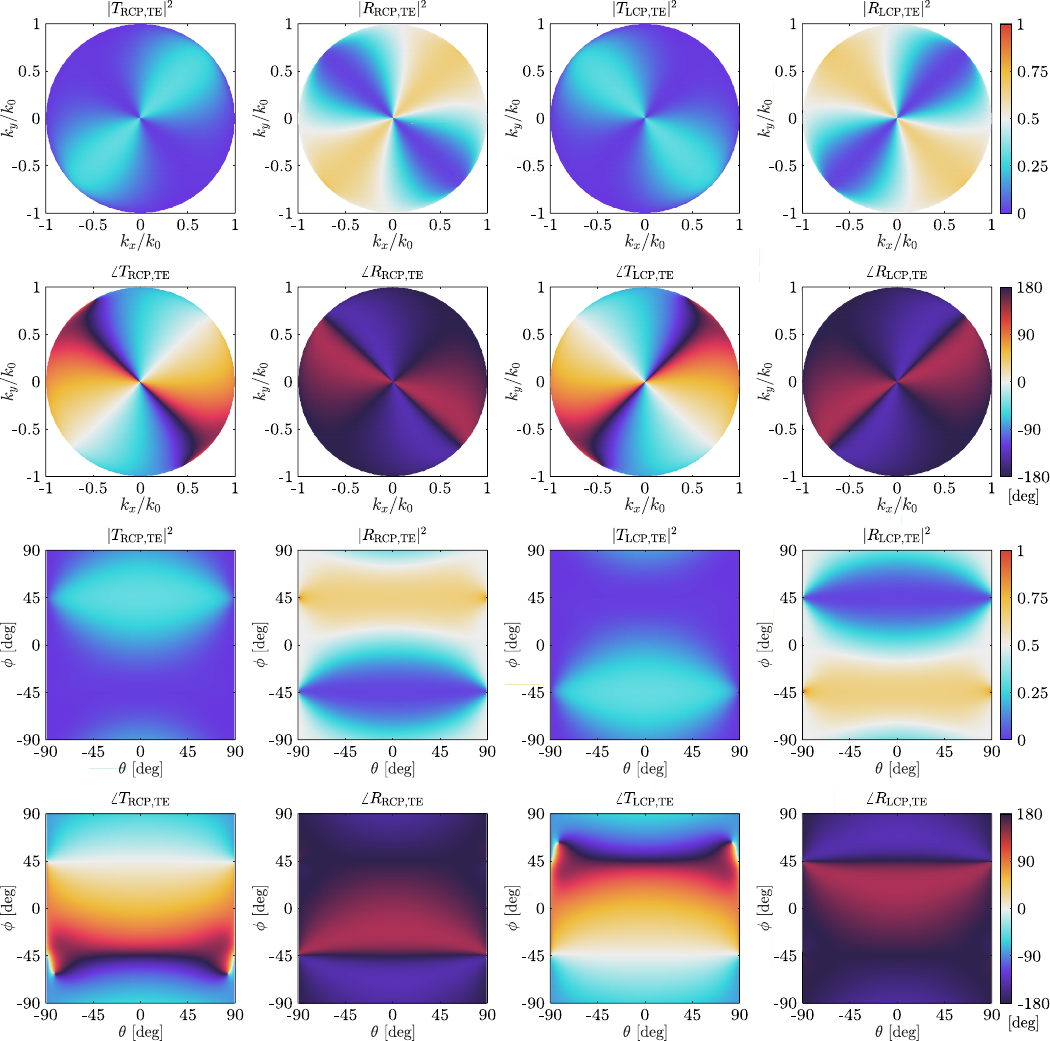}
\caption{Predicted results for linear-to-circular polarization, in the case of TE illumination. The top plots show the scattering in the $k_x,k_y$ plane; while the bottom plots display the scattering in terms of spherical coordinates angles ($\theta,\phi$).}\label{fig:mod_zz_pattern_cplp_te}
\end{figure}

\newpage

\begin{figure}[!h]
\centering
\includegraphics[width=1\linewidth]{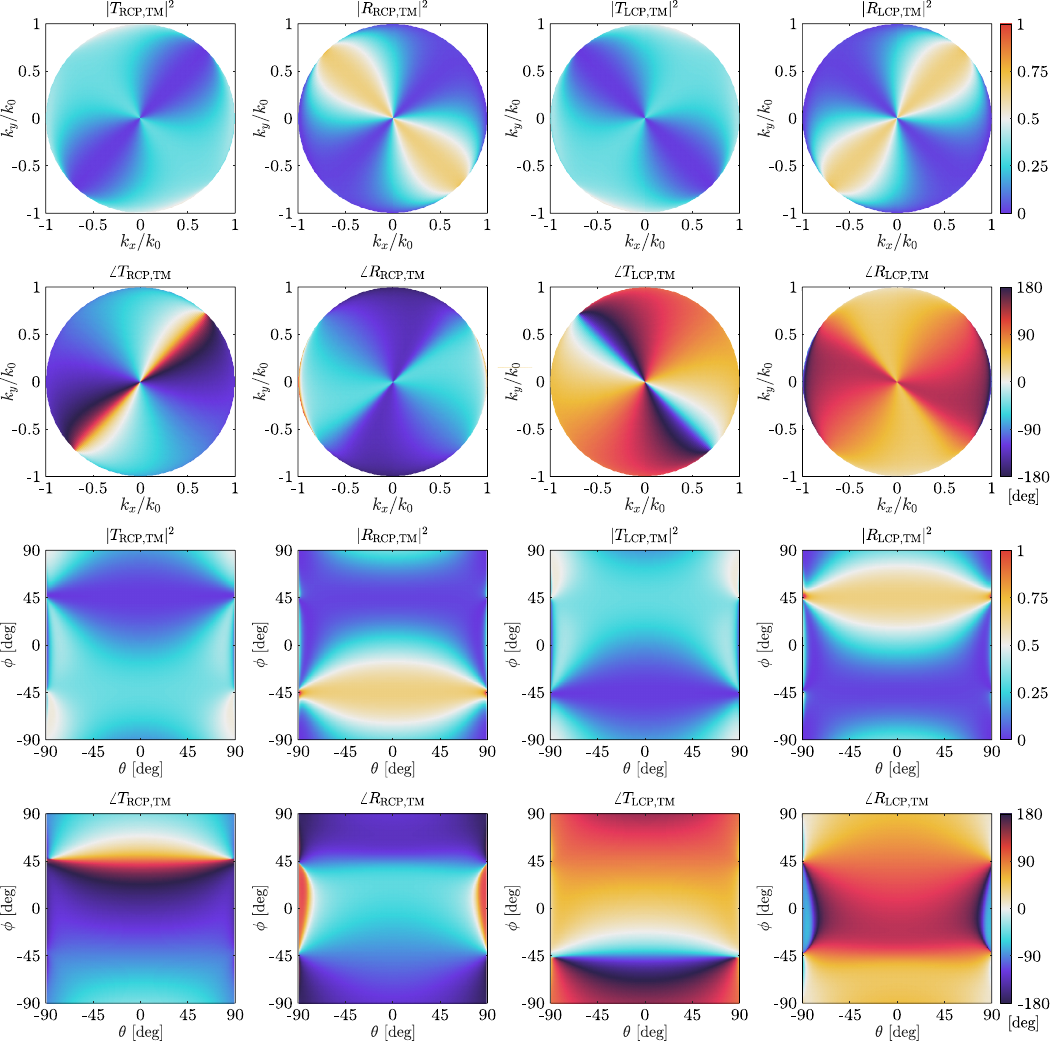}
\caption{Predicted results for linear-to-circular polarization, in the case of TM illumination. The top plots show the scattering in the $k_x,k_y$ plane; while the bottom plots display the scattering in terms of spherical coordinates angles ($\theta,\phi$).}\label{fig:mod_zz_pattern_cplp_tm}
\end{figure}

\end{document}